# Sustainable Pre-reduction of Ferromanganese Oxides with Hydrogen: Heating Rate-Dependent Reduction Pathways and Microstructure Evolution

Anurag Bajpai[1,*], Barak Ratzker[1], Shiv Shankar[1], Dierk Raabe[1,*], Yan Ma[1,2*]

[1]Max Planck Institute for Sustainable Materials, Max-Planck-Str. 1, Düsseldorf, 40237 Germany

[2]Department of Materials Science and Engineering, Delft University of Technology, Mekelweg 2, 2628 CD Delft, the Netherlands

*Corresponding authors: a.bajpai@mpie.de; d.raabe@mpie.de; y.m.ma@tudelft.nl

## Abstract

The reduction of ferromanganese ores into metallic feedstock is an energy-intensive process (2000-3000 kWh/ton) with substantial carbon emissions (1-1.5 tons $CO_2$ per ton metal), necessitating sustainable alternatives. Hydrogen-based pre-reduction of manganese-rich ores offers a low-emission pathway to augment subsequent thermic Fe-Mn alloy production. However, reduction dynamics and microstructure evolution under varying thermal conditions remain poorly understood. This study investigates the influence of heating rate (2-20 °C/min) on the hydrogen-based direct reduction of natural Nchwaning ferromanganese ore and a synthetic $Mn_2O_3$-9 wt.% $Fe_2O_3$ analog. Non-isothermal thermogravimetric analysis revealed a complex multistep reduction process with overlapping kinetic regimes. Isoconversional kinetic analysis showed increased activation energy with reduction degree, indicating a transition from surface-reaction to diffusion-controlled reduction mechanisms. Interrupted X-ray diffraction experiments suggested that slow heating enables complete conversion to MnO and metallic Fe, while rapid heating promotes Fe- and Mn-oxides intermixing. Thermodynamic calculations for the Fe-Mn-O system predicted the equilibrium phase evolution, indicating Mn stabilized Fe-containing spinel and halite phases. Microstructural analysis revealed that slow heating rate yields fine and dispersed Fe particles in a porous MnO matrix, while fast heating leads to sporadic Fe-rich agglomerates. These findings suggest heating rate as a critical parameter governing reduction pathway, phase distribution, and microstructure evolution, thus offering key insights for optimizing hydrogen-based pre-reduction strategies towards more efficient and sustainable ferromanganese production.

# 1. Introduction

The global effort to decarbonize industrial processes has placed hydrogen-based metallurgy at the forefront of sustainable metal production strategies [1, 2]. Among transition metals, manganese (Mn) plays a pivotal role in the production of high-strength steels, non-ferrous alloys, and energy storage materials [3]. Yet, Mn extraction remains a carbon-intensive metallurgical process. Conventionally, ferromanganese is produced through the carbothermic reduction of Mn ores in submerged arc furnaces, typically using coke as a reducing agent, with energy demands of 2000–3000 kWh/ton and greenhouse gas emissions exceeding 1–1.5 tons of $CO_2$ per ton of metal [4, 5]. The high process temperature and dependency on fossil carbon make the Mn production incompatible with emerging net-zero frameworks.

Hydrogen-based direct reduction (HyDR) offers a carbon-free alternative, replacing traditional reductants with hydrogen and producing only water vapor as a by-product [6]. Thermodynamically, hydrogen can reduce manganese oxides through a sequence of redox steps ($MnO_2 \rightarrow Mn_2O_3 \rightarrow Mn_3O_4 \rightarrow MnO$) culminating in the formation of MnO, which is thermodynamically stable under attainable $H_2/H_2O$ partial pressure ratios (~$10^3$) [7-9]. Although metallic Mn cannot be directly obtained *via* $H_2$ reduction due to its high affinity for oxygen, a two-step hybrid strategy has been proposed: a first stage of hydrogen-based pre-reduction to MnO, followed by a second-stage metallothermic extraction, typically using aluminum or carbon [10, 11]. This pre-reduction pathway is particularly attractive for low-carbon ferromanganese production, as it significantly reduces the oxygen load and subsequent reductant requirement [12]. Moreover, it can be carried out at moderate temperatures (600-900 °C), thereby offering improved energy efficiency [13]. However, this hybrid process introduces complexities. Natural ferromanganese ores often contain not only Mn oxides in various oxidation states but also 5–10 wt.% $Fe_2O_3$ and non-reducible gangue phases like silicates, Ca- and Al-oxides. Under HyDR conditions, $Fe_2O_3$ is readily reduced to metallic Fe through $Fe_3O_4$ and FeO intermediates [14, 15], whereas Mn oxides are only reduced to MnO [16]. Recent studies confirmed that under a 100% $H_2$ atmosphere, Fe oxides in Mn ore were completely converted to metallic Fe at 700-900 °C and exsolved as Fe particles, whereas MnO remained a stable matrix [13].

The formation of metallic Fe during the pre-reduction stage is not a passive byproduct, as the Fe morphology and distribution critically influence the chemical interactions with the MnO/Mn matrix during the downstream extraction step. In aluminothermic processes, for example, metallic Fe can consume aluminum *via* intermetallic formation, thereby reducing the availability of reductant for MnO conversion [17]. Hence, understanding and controlling the microstructure of pre-reduced Mn–Fe oxide systems is central to the success of hydrogen-assisted Mn extraction technologies.

Despite extensive literature on the $H_2$ reduction of individual Fe and Mn oxides, the reduction dynamics of mixed Mn–Fe oxides under hydrogen remain poorly understood. Most studies to date have focused on isothermal reduction kinetics of individual oxides or bulk ores [13], neglecting the fact that industrial reactors operate under non-isothermal conditions. Preliminary evidence from iron oxide reduction systems suggests that heating rate can significantly influence the reduction sequence, phase stability, and pore morphology [18]. Specifically, rapid heating may bypass intermediate oxide phases or cause kinetic overlap between reactions, while slow heating allows near-equilibrium progression of redox steps. For Mn-Fe systems, such thermal profiles may modulate the sequence and extent of Fe and Mn reduction, stabilize or suppress mixed oxide phases, and influence the exsolution and agglomeration behavior of metallic Fe. Yet, there have been no systematic investigations to elucidate how heating rate governs these phenomena in ferromanganese ores.

To address this critical knowledge gap, we performed a detailed investigation on the heating-rate-dependent hydrogen-based pre-reduction of ferromanganese oxides, using the Nchwaning ferromanganese ore, which contains ~9 wt.% $Fe_2O_3$. We applied non-isothermal thermogravimetric analysis (TGA) over a wide range of heating rates (2-20 °C/min) to uncover the kinetic regimes and the evolving activation energy profiles to delineate reaction mechanism transitions. To resolve the transient phase evolution, we conducted interrupted X-ray diffraction (XRD) experiments at specific reduction degrees and correlated them with thermodynamic predictions. Furthermore, we analyzed the microstructure evolution using scanning electron microscopy (SEM), energy-dispersive X-ray spectroscopy (EDS), and electron backscatter diffraction (EBSD), including detailed mapping of Fe and MnO grain structures and crystallographic orientation. These insights allowed us to directly link heating rate to metallic Fe particle morphology, MnO grain size, and residual phase content. Importantly, we discuss how these thermal-history-induced microstructural variations can influence the subsequent aluminothermic and carbothermic reduction behavior, with potential implications for Fe-Al intermetallic formation, metal-slag separation, and Mn yield. The findings from this study provide a robust mechanistic foundation for designing tailored pre-reduction strategies that optimize both energy efficiency and process yield in sustainable Mn metallurgy.

## 2. Methods

*2.1 Hydrogen-based direct reduction experiments*

Nchwaning ferromanganese ore (originally from the Kalahari manganese field in South Africa) was supplied by Oravské Ferozliatinové Závody (OFZ), Slovakia. A representative chemical analysis of the ore reveals approximately 48–53 wt.% Mn, 5–8 wt.% Fe, 4–7 wt.% $SiO_2$, and 4–7 wt.% CaO, with minor constituents including MgO (≤1 wt.%), $Al_2O_3$ (≤0.5 wt.%). The dominant Mn-bearing phase is bixbyite ($Mn_2O_3$: 78.5±4.6 wt.%) with small amounts of braunite (($Mn^{2+}Mn^{3+})_6(SiO_4)O_8$: 2.6±1.8 wt.%). The iron present in the ore predominantly exists in the form of hematite ($Fe_2O_3$: 7.7±3.3 wt.%) [13]. Roughly disc-shaped samples, approximately 15 mm in diameter and weighing between 1.2 and 1.4 g, were cut from the ore chunks using a diamond wire saw. For comparative studies, synthetic samples of similar $Mn_2O_3$:$Fe_2O_3$ stoichiometry to the Nchwaning ore were prepared by mixing high-purity (97-98%) oxide powders (Sigma Aldrich). For this, 91 wt.% $Mn_2O_3$ powder was mixed with 9 wt.% $Fe_2O_3$ powder, followed by mechanical milling at 250 rpm for 6 h, where simultaneous mixing and reduction in particle size yielded a microscale oxide powder mixture. Thereafter, the powder mixture was cold compressed into pellets of 15 mm diameter and 1.2-1.4 g weight. The elemental composition of the ore and synthetic samples can be found in Table S1.

HyDR of the samples was carried out in an in-house thermogravimetric analysis (TGA) setup in a vertical quartz tube furnace [19]. Temperature-programmed experiments were conducted according to the following protocol, with the temperature being measured and controlled by a thermocouple inserted into the center of a reference pellet. For each reduction experiment, a sample was laid on a 304 stainless steel mesh in an open quartz basket connected to a high-precision balance (0.1 µg resolution) and exposed to hydrogen (99.99% purity) gas at a flow rate of 10 L/h. The sample was heated using infrared light to 900 °C at heating rates of 2, 5, 10 and 20 °C/min and then held isothermally at 900 °C for 10 min. After the isothermal period, the furnace was powered down, allowing the samples to cool rapidly within the furnace. The mass change of each sample was continuously monitored and recorded during the reduction process. The degree of reduction was calculated based on the experimental mass loss divided by the theoretical mass loss, considering a complete reduction of $Mn_2O_3$ and $Fe_2O_3$ into metallic Mn and Fe, respectively.

*2.2 Phase analysis and Microstructure characterization*

For microstructure analysis, a disk sample approximately 1 mm thick was prepared from the cross-section of the samples. Metallographic preparation of the samples included mounting in a conductive resin and grinding with SiC abrasive papers, followed by polishing using diamond microparticles and silica nanoparticle suspensions. The phases of the initial and reduced samples were identified and quantified by X-ray diffraction (XRD) using a RIKAKU SMARTLAB 9KW diffractometer with Cu K$_\alpha$ radiation ($\lambda$ = 1.54059 Å). The scanning range 2$\theta$ was from 20° to 120° with a scanning step of 0.01° and a scanning speed of 2° min$^{-1}$. The local microstructures of the samples were characterized using scanning electron microscopy (SEM) coupled with electron backscatter diffraction (EBSD) for local phase and crystallography analysis and energy dispersive spectroscopy (EDS) for local chemical analysis. A Zeiss Sigma scanning electron microscope was used to examine the microstructures, operated at an accelerating voltage of 15 kV and a current of 7 nA. EBSD maps were collected using the EDAX DigiView 5 camera and EDAX APEX software. The Kikuchi patterns were indexed by the spherical indexing method using the EDAX OIM Matrix v.9 software [20]. The pore characteristics of the initial and reduced samples were quantitatively analyzed using scanning electron microscopy (SEM). SEM images were acquired at 2000× magnification. Image processing was carried out using custom Python scripts, where grayscale thresholds were manually tuned to segment pore regions from the solid matrix with high fidelity. For each sample, 5 micrographs were analyzed to obtain statistically representative porosity data. The porosity was calculated as the area fraction of segmented pores relative to the total image area, averaged over all images. To estimate pore size, the equivalent circular diameter ($d_{eq}$) was determined using: $d_{eq} = 2 \times \sqrt{A/\pi}$, where $A$ is the projected area of each pore.

*2.3 Thermodynamic calculations*

Equilibrium calculations for the Fe-Mn-O system based on the CALculation of PHAse Diagram (CALPHAD) approach were performed using the ThermoCalc software version 2024a. The thermodynamic calculations were performed using the metal oxide solutions TCOX10 database and the equilibrium state is calculated by the Gibbs energy minimization of the system. The input composition was based on a mixture of Fe$_2$O$_3$ (9 wt.%) and Mn$_2$O$_3$ (91 wt.%), and the corresponding number of moles of Fe, Mn, and O atoms was used for equilibrium calculations at 400 and 600 °C. To estimate the oxygen partial pressure (pO$_2$) in the system, the total number of oxygen atoms was systematically varied from zero (representing the fully metallic state) up to the stoichiometric oxide state. The effects of temperature and pO$_2$ on the phase evolution were investigated. Additionally, the elemental mole fractions within individual phases were calculated as a function of pO$_2$ to gain insight into elemental partitioning during the reduction reactions.

## 3. Results

### *3.1 Initial microstructures*

Figure 1a shows backscattered electron (BSE) images and EDS maps for the natural Nchwaning ore. The ore is relatively dense (~11% porosity) with mainly isolated pores. Figure S1 provides the distribution of the equivalent pore size for the Nchwaning ore sample. It comprises mostly angular grains with relatively smaller equiaxed second phase grains with noticeable phase contrast in the BSE image (Figure 1a). EDS maps (top row) display microscale heterogeneity with discrete Fe-rich oxide particles (red) embedded in a Mn-oxide matrix (green), with additional minor Ca-, Si-bearing gangue phases (Figure S2). Given the abundant amount of Fe-rich phase (~ 6 wt.%), it is imperative to elucidate the mechanistic aspects of its simultaneous reduction with the Mn-oxide matrix during HyDR. However, gangue in the ore can be detrimental to a systematic investigation of the interplay between $Mn_2O_3$ and $Fe_2O_3$: silicate and oxide phases can melt or react at high temperature (≥ 900 °C) to form viscous Mn-silicates, effectively "locking up" Mn as silicate and hindering its reduction to MnO [21]. Therefore, a synthetic analog, mimicking the stoichiometric proportions of $Mn_2O_3$ and $Fe_2O_3$ from the ore (91 wt.% $Mn_2O_3$:9 wt.% $Fe_2O_3$), was used to understand the reaction mechanisms involving Fe and Mn mixed oxides.

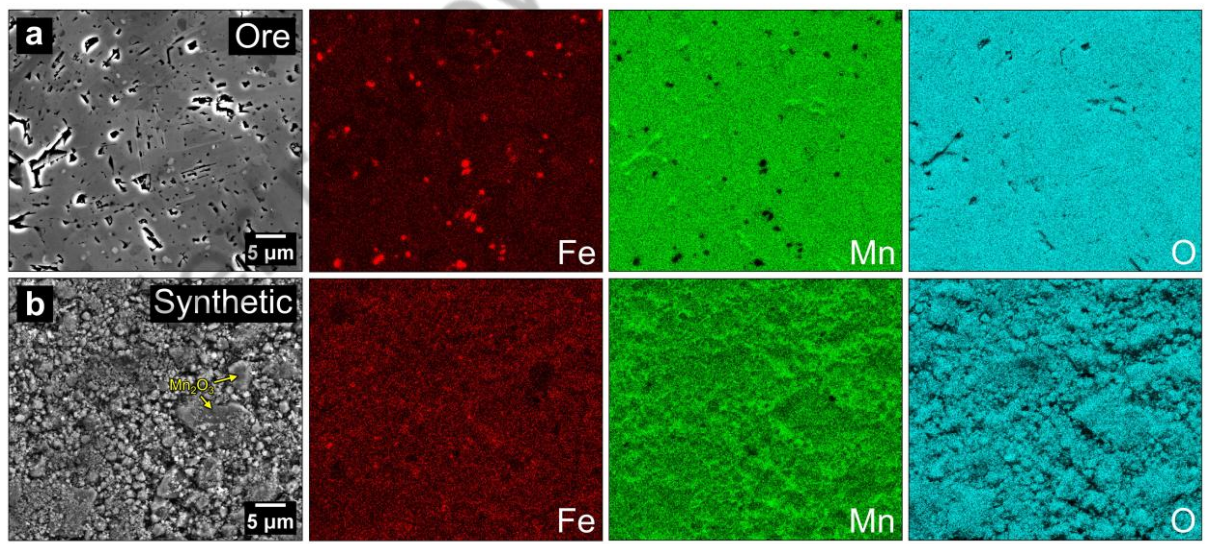

**Figure 1 -** Microstructural and compositional comparison of (a) Nchwaning ore and (b) $Mn_2O_3$-9$Fe_2O_3$ (wt.%) synthetic powder. Elemental EDS maps for Fe (red), Mn (green), and O (cyan) demonstrate pronounced spatial heterogeneity in the ore, particularly with localized Fe enrichment and Mn segregation, while the synthetic sample shows a more uniform elemental distribution.

Figure 1b shows the BSE images and EDS maps for the synthetic analog sample. In contrast to the ore, the synthetic sample is completely porous and encompasses roughly spherical particles. Although some relatively large few-micron sized $Mn_2O_3$ particles can be observed, most of the sample consists of fine mixture of submicron particles. The EDS mapping shows a homogeneous intermixture of Mn and Fe signals, suggesting a homogeneous dispersion of $Fe_2O_3$ among $Mn_2O_3$. Higher porosity (due to cold compaction of micron-sized spherical particles) and homogeneity allow for easy and uniform gas exchange during reduction and circumvent the impact of gradual topochemical reduction [22, 23] and chemical heterogeneity observed in natural ores [24]. In essence, the ore microstructure is dense, coarse, and multiphase ($Mn_2O_3$/$Fe_2O_3$ grains plus silicate/carbonate gangue), whereas the synthetic compact is a finely mixed Mn-Fe oxide matrix without any impurities, ideal for decoupling the stepwise reduction reactions between $Mn_2O_3$ and $Fe_2O_3$.

*3.2 Reduction kinetics*

Figure 2a shows the plot for conversion degree ($\alpha$) *vs.* temperature ($T$) of the Nchwaning ore sample, revealing complex yet discernible multistep reaction reduction profiles. For all heating rates, there are two pronounced sequential mass-loss stages. The $\alpha$–$T$ curve for the ore reduced at the slowest heating rate (2 °C/min) shows negligible weight loss below ~300 °C, a minor first step at ~350-450 °C, and a dominant second step at ~500-750 °C. Increasing the heating rate ($\beta$ = 2→20 °C/min) shifts both steps to higher temperature (by ~50-100 °C) and sharpens the $\alpha$–$T$ profiles, consistent with thermal lag. The total mass loss observed for the ore is on the order of ~10-12%, consistent with nearly complete reduction of its Mn- and Fe-oxides to lower oxide (MnO) and Fe metal (with the theoretical mass loss of 11.9%). In particular, the high-temperature plateau (above 800 °C) in the conversion curve suggests that by ~900 °C the reduction was halted with a maximum conversion degree of 38.56% (~98.7% of the theoretical value, Table 1), which corresponds to nearly complete reduction of $Mn_2O_3$ to MnO and $Fe_2O_3$ to metallic Fe. The multistep nature of the reduction can be clearly discerned by the reduction rate (d$\alpha$/d$t$) *vs.* temperature ($T$) plot, shown in Figure 2c, demonstrating multiple distinct peaks for each $\beta$ in the reduction rate. At the lowest heating rate (2 °C/min), these peaks are more distinctly resolved, whereas at higher rates (20 °C/min), they shift to higher temperatures and overlap. This fact reflects the effect of heating rate: slow heating allows individual reduction steps to occur stepwise, while fast heating compresses and overlaps the events.

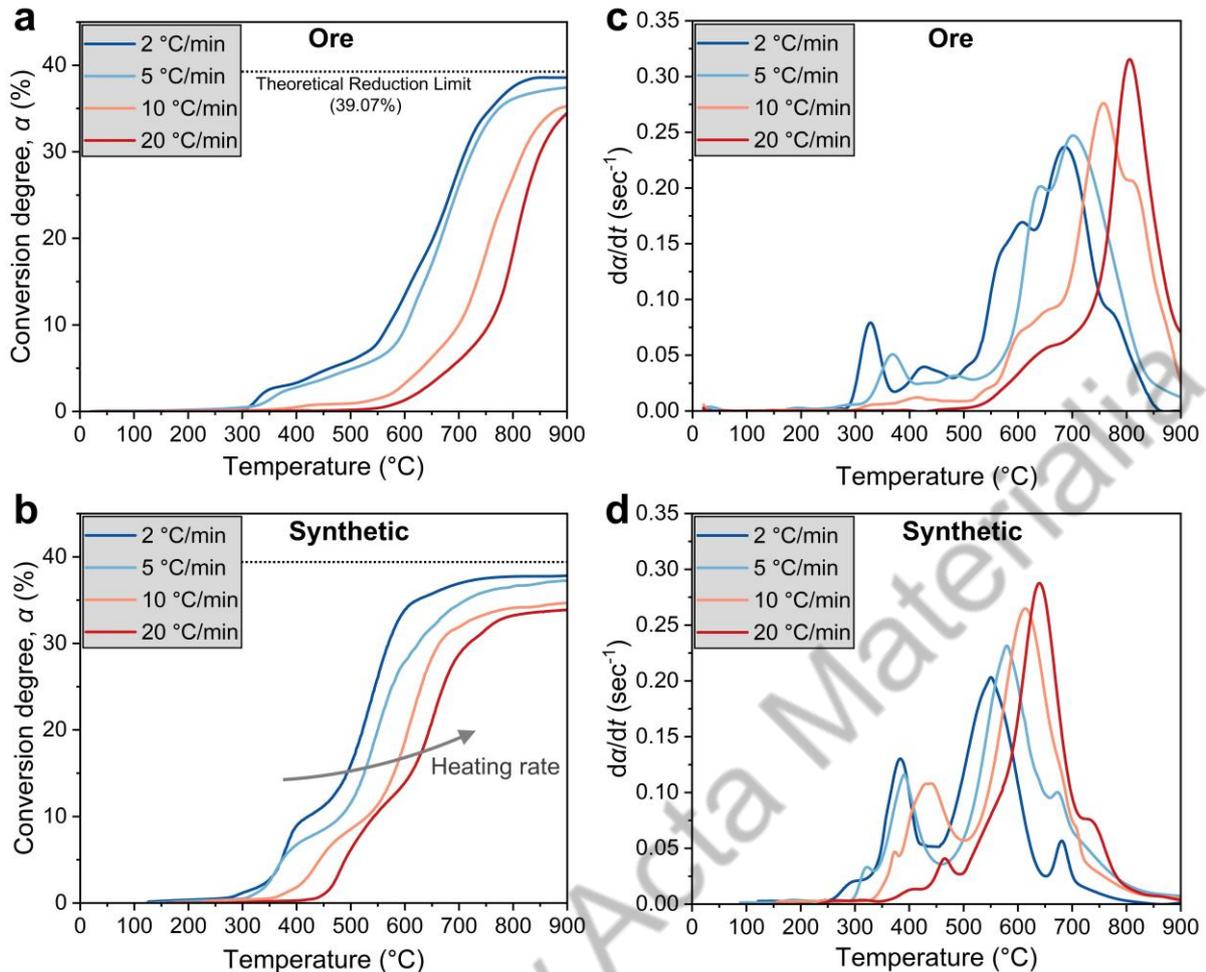

**Figure 2 -** Non-isothermal conversion degree as a function of temperature during the hydrogen-based direct reduction of (a) Nchwaning ore and (b) $Mn_2O_3$-9$Fe_2O_3$ (wt. %) compact synthetic analog at different heating rates; Corresponding reduction rate (d$\alpha$/d$t$ vs. $T$) for (c) Nchwaning ore and (d) synthetic samples, revealing the presence of multiple overlapping kinetic events, with distinct maxima for each heating rate.

The synthetic analog sample shows a broadly similar multistage reduction behavior (Figure 2b), but with more obvious changes in reduction rate, corresponding to different reduction reactions. At 2 °C/min, the final conversion degree reaches 37.82%, closely approaching the theoretical maximum (Table 1) and confirming that the sample is nearly fully reduced to MnO and Fe by ~900 °C. However, it is noteworthy that at lower heating rates (2-5 °C/min), the synthetic sample shows slightly lower final conversion than the ore. This behavior can be attributed to potentially early sintering and densification in the cold-pressed fine-powder compact, which may reduce gas permeability and inhibit complete reduction of the particle interiors. In contrast, the coarse-grained structure and intrinsic intergranular porosity in the ore promote more uniform reduction under slow heating. As the heating rate increases, sintering is suppressed in both materials, and the final conversion degrees converge. Figure 2d shows the

d$\alpha$/d$t$–$T$ plot for the synthetic sample, showing more well-resolved peaks, indicating a more distinct multistep reduction behavior of the synthetic analog compared with the natural ore. As with the ore, higher heating rates shift these peaks to higher temperatures and increase their overlap. Despite the broader transitions in the ore, the similarity in overall $\alpha$–$T$ profiles between the natural and synthetic samples suggests both cases undergo a comparable sequence of redox reactions. This indicates that the underlying mechanistic aspects of the simultaneous reduction of $Mn_2O_3$ and $Fe_2O_3$ are fundamentally shared between the ore and the synthetic analog systems.

**Table 1 –** Conversion degree (and fraction over the theoretical value) as a function of heating rate for Nchwaning ore and synthetic sample reduced at 900 °C (and held isothermally for 10 min).

|  |  | Theoretical value** | 2 °C/min | 5 °C/min | 10 °C/min | 20 °C/min |
|---|---|---|---|---|---|---|
| **Conversion Degree (%)** | **Nchwaning Ore*** | 39.07 | 38.56 (98.7%) | 37.83 (96.8%) | 35.77 (91.5%) | 34.86 (89.2%) |
|  | **Synthetic Analog** | 39.07 | 37.82 (96.8%) | 37.23 (95.3%) | 36.01 (92.1%) | 35.18 (90.1%) |

\* The ore conversion values are based on the known stoichiometry of $Mn_2O_3$ and $Fe_2O_3$, excluding contributions from gangue components.
\*\* The theoretical conversion degree here assumes that $Mn_2O_3$ is reduced to MnO and $Fe_2O_3$ is reduced to metallic Fe.

*3.3 Activation energy analysis*

Isoconversional model-free kinetic analysis, *i.e.*, Kissinger–Akahira–Sunose (KAS) [25], Ozawa–Flynn–Wall (OFW) [26], and the differential Friedman method [27], were applied to the non-isothermal TGA data to determine the apparent activation energy ($E_a$) as a function of conversion degree ($\alpha$) for each sample (more information on the three isoconversional methods in Supplementary Note S1). Figure S4 details the linear-fit plots used in the KAS, OFW, and Friedman analyses for different $\alpha$ levels (5–35%). All three isoconversional methods yielded consistent linear trends across the evaluated conversion range (5–35%). The near linearity of these plots for all conversion levels supports the validity of the Arrhenius assumption within $\alpha$ intervals. Minor deviations and increased scatter were observed at higher conversion levels ($\alpha \geq 30\%$), particularly in the differential (Friedman) analysis. KAS and OFW, being integral methods, yield smoother $E_a(\alpha)$ profiles, while the Friedman differential method is sensitive to experimental noise and gives slightly higher values (than KAS and OFW) [27]. Thus, KAS and OFW methods reveal the conversion dependence, and Friedman provides details on subtle changes in the kinetic data [25, 28]. Nevertheless, the agreement among these three methods on the $E_a(\alpha)$ builds confidence in the interpretation that HyDR proceeds *via* overlapping sub-reactions, each reaction with its own rate-controlling step and energy barrier [29].

The resulting $E_a(\alpha)$–$\alpha$ profiles are shown in Figure 3a and b for the ore and synthetic analog, respectively (and provided in Table S2). Both samples display a progressive increase in the $E_a(\alpha)$ with the progression of reduction. The upward trend in $E_a(\alpha)$ indicates that the last stages of reduction are kinetically more demanding than the initial stages. At low conversions (~5-20%), the apparent $E_a(\alpha)$ is relatively low (~10-25 kJ/mol), but $E_a(\alpha)$ rises steadily as the reaction proceeds. For the ore, $E_a(\alpha)$ is around ~15-25 kJ/mol in the early stage and increases to ~50-70 kJ/mol (from the KAS method) by 35% conversion. These values are in line with previous reports for $Mn_3O_4 \rightarrow MnO$ reduction (~50-110 kJ/mol) [10] and hematite reduction (~50-80 kJ/mol) [30]. The synthetic sample shows a similar trend, with $E_a(\alpha)$ increasing from ~15-20 kJ/mol to ~60-80 kJ/mol over the 5-35% conversion range. The non-constant $E_a(\alpha)$ behavior also confirms that the reduction cannot be described by a single activation energy; instead, the controlling mechanism changes as the reaction progresses from an initially fast, chemical-reaction (or nucleation) regime to a slower, diffusion-controlled regime [10, 31].

Importantly, the ore sample shows consistently higher activation energies than the synthetic analog in the early stages ($\alpha$ = 5-20%), which is attributed to the coarser grain size and lower surface area of the natural ore, limiting initial gas-solid interactions. In contrast, the synthetic compact, with its fine particle size and higher surface area, shows faster initial reduction, reflected in lower $E_a$ values. However, as the reaction progresses beyond $\alpha \approx 20\%$, the trend reverses, and the synthetic sample exhibits higher $E_a$ than the ore, particularly evident from $\alpha$ = 30-35%. This shift likely arises from early-stage sintering and densification in the synthetic compact, which increases diffusional resistance during later-stage reduction. The natural ore, with its more stable porous network and microcrack-assisted transport pathways, maintains relatively lower barriers at higher conversion degrees [32].

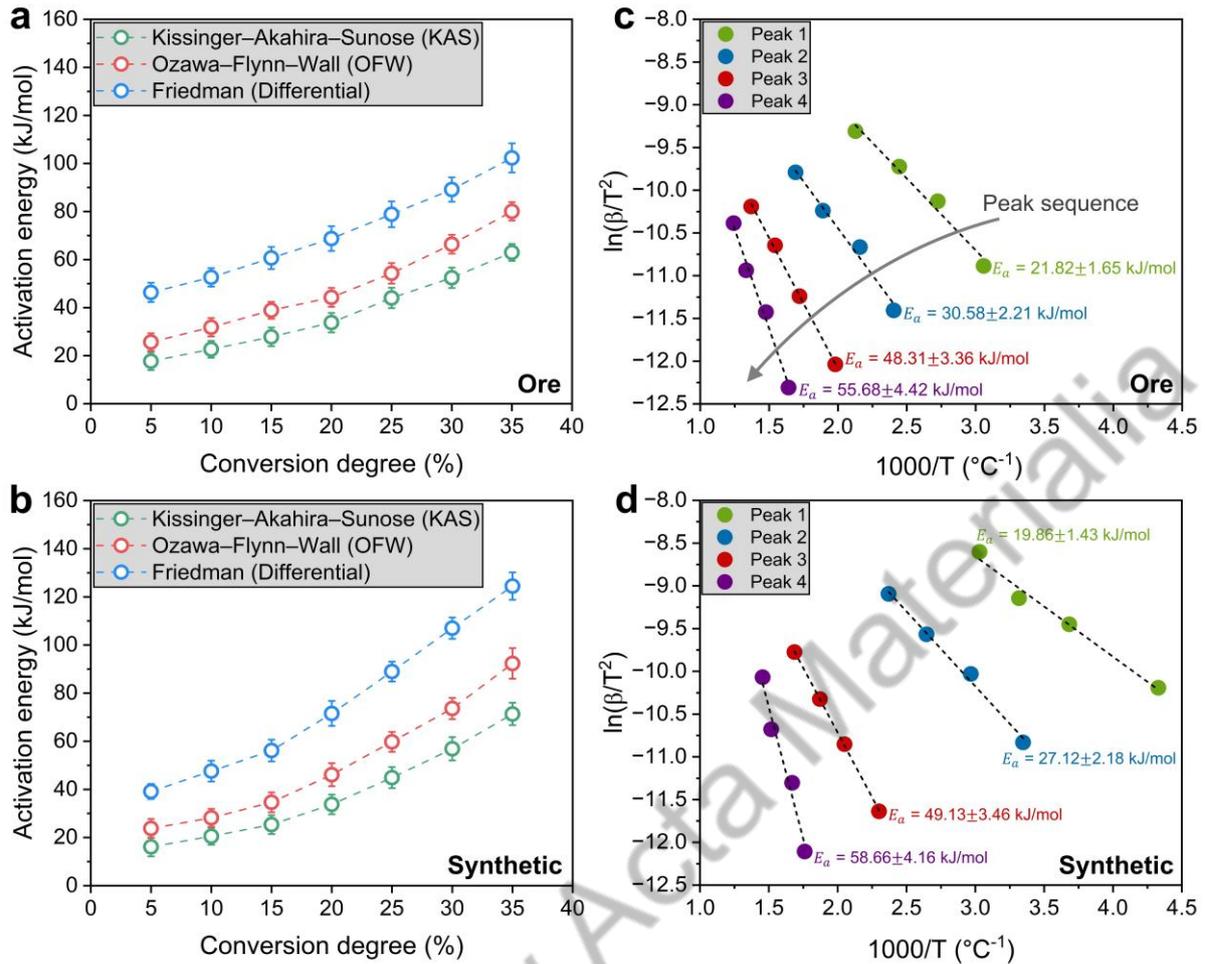

**Figure 3 -** Isoconversional activation energy ($E_a$) profiles determined using the Kissinger–Akahira–Sunose (KAS), Ozawa–Flynn–Wall (OFW), and Friedman methods for (a) Nchwaning ore and (b) $Mn_2O_3$-$9Fe_2O_3$ (wt.%) synthetic analog, showing a clear increase in the apparent activation energy with conversion degree, indicative of a multistep reaction reduction process with changing rate-controlling mechanisms. Kissinger's review of four distinct derivative peaks for (c) Nchwaning ore and (d) synthetic samples, the calculated activation energies are also indicated.

To complement the isoconversional approach, peak-based Kissinger analysis was further performed for each distinguishable peak in the d$\alpha$/d$t$-$\alpha$ plots (Figure 2c, d) for both samples. Arrhenius plots of $\ln(\beta/T_p^2)$ vs $1000/T_p$ for four peaks (Figure 3c, d) yielded apparent activation energies ranging roughly from ~20 kJ/mol to ~60 kJ/mol for the individual reduction events. For the ore, the lowest-temperature peak gave $E_a$≈22 kJ/mol, while the highest-temperature peak gave $E_a$≈56 kJ/mol (Figure 3c). The synthetic sample showed a similar range ($E_a$≈20 to 59 kJ/mol for its four peaks, Figure 3d). While the synthetic sample exhibits lower activation energies than the ore for the initial reduction events (Peaks 1 and 2), reflecting its finer particle size and faster early-stage kinetics, this trend reverses at higher temperatures. For Peaks 3 and 4, the synthetic compact shows higher $E_a$ values than the ore, indicative of potential sintering-induced diffusion limitations that may emerge during the later stages of reduction. Further, the overall $E_a$ values are lower than typical literature values for single Mn-oxide reductions with various morphologies (in the range of 50–120 kJ/mol) [33]. This discrepancy likely reflects the apparent, averaged energy barriers of overlapping sub-reactions rather than distinct single-step transitions. Nonetheless, the multi-peak structure and systematic evolution of $E_a$ values confirm that hydrogen-based reduction of Mn–Fe oxides proceeds through a sequence of overlapping redox events with shifting rate-controlling steps.

### *3.4 Phase evolution*

The phase composition of the ferromanganese Nchwaning ore and synthetic sample before and after HyDR (at the heating rate of 10 °C/min) was determined by XRD. The ore sample exhibits a complex mineralogy, with Mn (III) oxide and multiple gangue phases, as shown in Figure 4a. Specifically, XRD peaks correspond to Mn (III) oxide (bixbyite, $Mn_2O_3$), braunite I (mixed $Mn^{2+}$–$Mn^{3+}$ silicate phase), and hausmannite ($Mn_3O_4$), along with minor hematite ($Fe_2O_3$) and other gangue (like oxides, calcites, and silicates). The peaks related to gangue phases are given in detail in Figure S5, with a zoomed-in view (20-40° 2θ region) of the XRD pattern. After HyDR, this multiphase assemblage collapses predominantly into mainly two phases, namely MnO (manganosite) and $\alpha$-Fe (metallic Fe in body-centered cubic structure). This indicates that the Mn(III) oxide in the ore is largely converted to MnO, and $Fe_2O_3$ is reduced to metallic Fe. Figure 4b shows XRD patterns for the synthetic analog sample before and after HyDR. It confirms that the pressed pellet consisted of bixbyite $Mn_2O_3$ and ~9 wt.% hematite $Fe_2O_3$ initially, and they were reduced to MnO and metallic Fe, respectively, during HyDR.

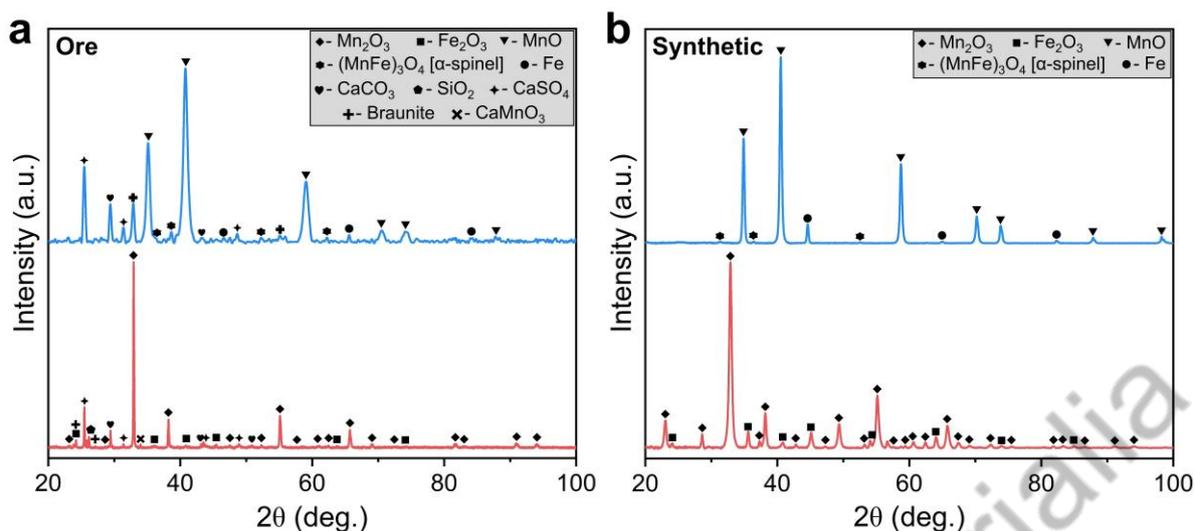

**Figure 4 -** XRD patterns of (a) Nchwaning ore and (b) $Mn_2O_3$-$9Fe_2O_3$ (wt.%) synthetic analog before (red) and after (blue) hydrogen-based direct reduction at 900 °C with a heating rate of 10 °C/min.

Figure 5a presents the XRD patterns of the synthetic analog sample reduced at different heating rates. The heating rate exerts a subtle influence on the phase composition of the end product. At the lower heating rates (2 and 5 °C/min), the reduced synthetic sample possesses MnO and Fe, as confirmed by XRD and no other phases are detected. This result suggests essentially a complete reduction of $Mn_2O_3$ and $Fe_2O_3$ to MnO and Fe. At higher heating rates (10 and 20 °C/min), however, a trace of tetragonal $Mn_3O_4$ (hausmannite) is detected. The intensity of the residual tetragonal spinel ($Mn_3O_4$) peaks increases with heating rate (as shown in the magnified diffractogram). No peaks for Fe-oxides (such as $Fe_3O_4$ or FeO) are detected in any reduced samples, suggesting that Fe-oxide either reduces all the way to the metallic form or dissolves into Mn-rich oxide phases (as $Fe^{2+}/Fe^{3+}$ in $Mn_3O_4$ or MnO). Figure 5b shows the phase fractions in the reduced synthetic samples, as analyzed by Rietveld refinement. At a heating rate of 2 °C/min, the product consists of ~91.8±3.4 wt.% MnO and ~8.2±2.8 wt.% Fe. A nearly identical phase split is found at a heating rate of 5 °C/min (~92.0±3.8 wt.% MnO, 8.0±3.1 wt.% Fe). At a heating rate of 10 °C/min, however, about 2.2±1.6 wt.% of the product is $Mn_3O_4$ spinel, with ~90.9±4.1 wt.%MnO and ~6.9±2.2 wt.%Fe. The highest heating rate of 20 °C/min yields a similar phase composition as the sample reduced at the heating rate of 10 °C/min, i.e., ~2.8±1.4 wt.% $Mn_3O_4$, ~90.8±4.4 wt.% MnO and ~6.3±2.8 wt.% Fe. Within experimental uncertainty, the metallic Fe yield stays roughly constant (~6-8%) in all samples reduced at different heating rates, whereas the amount of unreduced tetragonal spinel increases from ~0 to ~3% as the heating rate rises from 2 to 20 °C/min. These findings suggest that while Fe reduction proceeds to completion regardless of thermal history, higher heating rates result in partial kinetic arrest of the Mn-oxide reduction, leading to residual hausmannite in the final product.

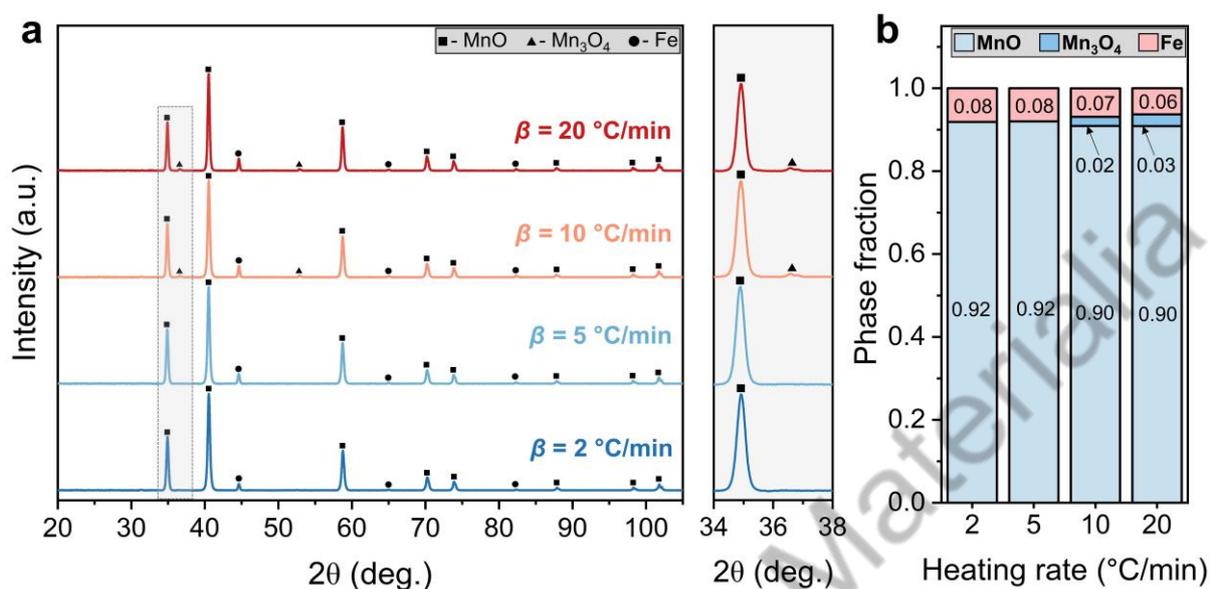

**Figure 5** – (a) XRD patterns of reduced $Mn_2O_3$-$9Fe_2O_3$ (wt.%) synthetic analog samples obtained after hydrogen-based reduction at different heating rates (2, 5, 10, and 20 °C/min). The side panel highlights the detectable $Mn_3O_4$ spinel peaks. (b) The calculated phase fractions for samples reduced at different heating rates.

## 3.5 Microstructure evolution

Figure 6 shows the microstructure of the ferromanganese Nchwaning ore (Figure 6a, b) and synthetic analog sample (Figure 6c, d) following hydrogen-based direct reduction (HyDR) at 900 °C under two representative heating rates: 5 °C/min and 20 °C/min. It should be noted that reduced samples at the heating rate of 2 and 5 °C/min display similar morphologies, the same as the samples at the heating rate of 10 and 20 °C/min, as shown in Figure S6. Therefore, the in-depth microstructure characterization was conducted for two samples reduced at the heating rates of 5 °C/min and 20 °C/min. Both materials exhibit a dual-phase microstructure comprising metallic Fe and MnO. However, their morphology, phase distribution, and porosity characteristics vary substantially with heating rate and between the natural ore and synthetic analog.

For the sample reduced at 5 °C/min, the BSE micrographs reveal a relatively fine and homogeneously distributed microstructure. In the ore sample (Figure 6a), Fe appears as uniformly dispersed sub-micron particles (<0.8 μm) finely embedded within a porous MnO matrix. The corresponding elemental maps confirm that the Fe phase is distributed across the MnO matrix without significant agglomeration. In the synthetic sample (Figure 6c), a similar interpenetrating morphology is observed, though with even finer dispersion. Owing to the homogeneous precursor mixture, the Fe precipitates in the synthetic analog are ultrafine (<0.3 μm) and distributed within the MnO framework.

At a higher heating rate of 20 °C/min, both systems exhibit markedly coarser and more heterogeneous microstructures. In the ore (Figure 6b), large Fe-rich regions (≥8 μm) appear as bright features, frequently forming encapsulation-like morphologies where metallic Fe surrounds residual MnO cores. These features suggest accelerated Fe coarsening and localized sintering during rapid heating. The MnO phase displays regions of varying density, ranging from dense islands to nanoporous domains. Similar trends are observed in the synthetic sample (Figure 6d), with Fe forming continuous clusters and wrapping around dense MnO cores. Nanoporous and dense MnO regions coexist, highlighting the heterogeneous evolution of microstructure at high thermal heating rates. The overall microstructure morphology thus transitions from a fine and interwoven Fe-MnO microstructure in the sample reduced at 2-5 °C/min to a dual-morphology microstructure with additionally coarse Fe-rich regions in the sample reduced at 10-20 °C/min. This aligns with literature on hydrogen-reduced Fe oxides, where slow reduction yields porous Fe with many fine pores, while fast heating produces dense Fe regions with large pores [34].

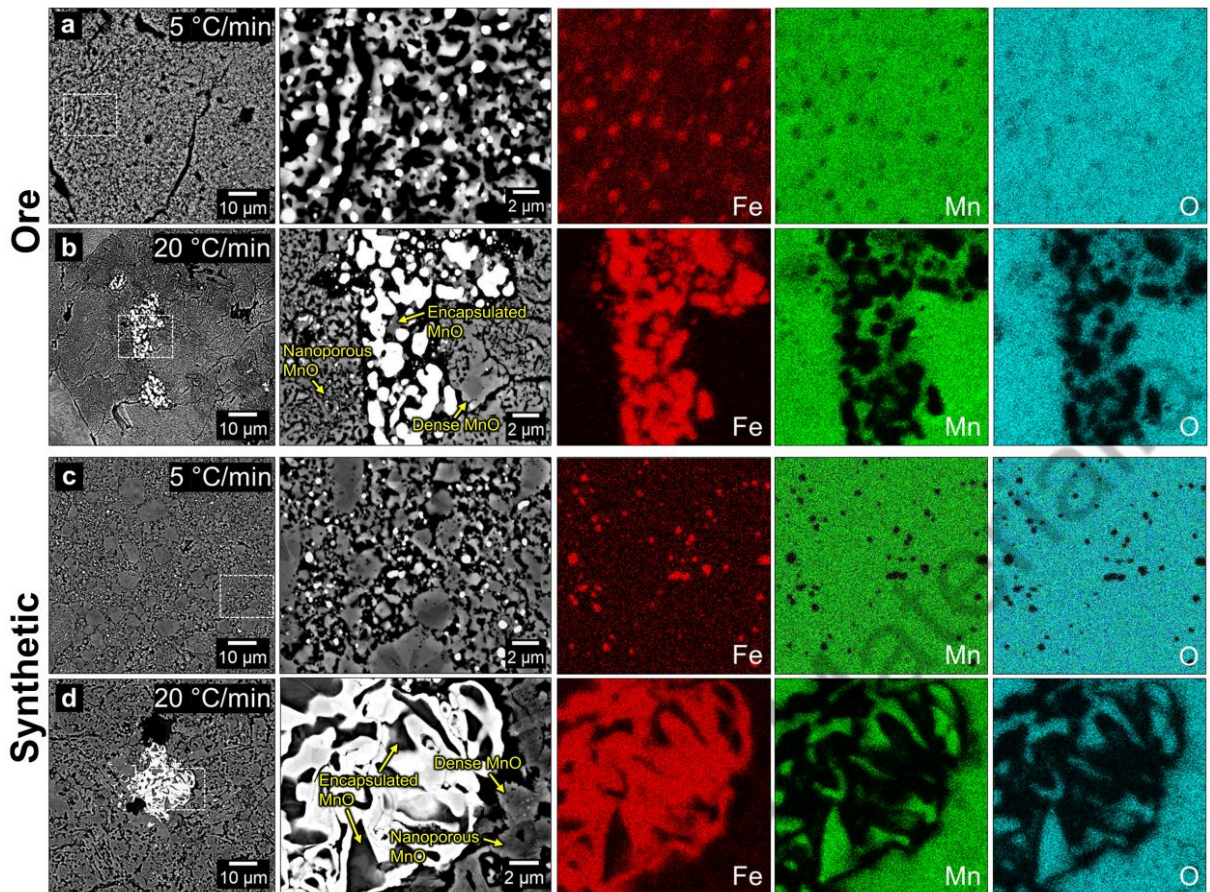

**Figure 6 -** Microstructure of ferromanganese reduced at 900°C; BSE images at low and high magnification alongside corresponding EDS mapping of Fe (red), Mn (green), and O (cyan). Nchwaning ore samples reduced with a heating rate of (a) 5 °C/min and (b) 20 °C/min, and $Mn_2O_3$-$9Fe_2O_3$ (wt.%) synthetic sample reduced with a heating rate of (c) 5 °C/min and (d) 20 °C/min.

The microstructure transitions align with changes in porosity and pore size distribution. Figure 7 quantifies the total porosity in both ore and synthetic analog samples as a function of heating rate. For both the ore and synthetic samples, porosity increases systematically with heating rate from ~15% at 2 °C/min to ~25-30% at 20 °C/min. However, when comparing the two materials, subtle differences emerge. At low heating rates (2 and 5 °C/min), both samples show comparable porosity values (~14-18%). In contrast, at higher heating rates (10 and 20 °C/min), the synthetic sample consistently exhibits higher porosity than the ore. This difference is attributed to its powder-derived morphology and higher initial surface area, which inhibit densification and promote pore growth during rapid heating. In contrast, the ore, being initially more compact and less porous (~11%), forms most of its interconnected porosity during the reduction process itself. Consequently, the synthetic sample exhibits greater sensitivity to heating rate, both in terms of porosity development and metallic Fe dispersion, reinforcing the influence of starting microstructure on reduction-induced morphological evolution.

The porosity morphological differences are further clarified through pore size distribution analysis (Figure S7). At low heating rates (2 and 5 °C/min), both samples display narrow distributions centered below ~1.0 μm, with slightly finer pores in the synthetic sample. As the heating rate increases, the distributions shift toward larger diameters for both systems. At 10 °C/min, the average pore size increases to ~1.6 μm for the ore and ~1.5 μm for the synthetic analog. At 20 °C/min, the average diameter rises further to ~2.1 μm for the ore and ~1.9 μm for the synthetic sample, though the synthetic sample still exhibits a slightly narrower distribution (lower standard deviation). The increase in pore size with heating rate reflects accelerated reaction front advancement and rapid gas evolution at high heating rates, which inhibit densification and promote pore coarsening.

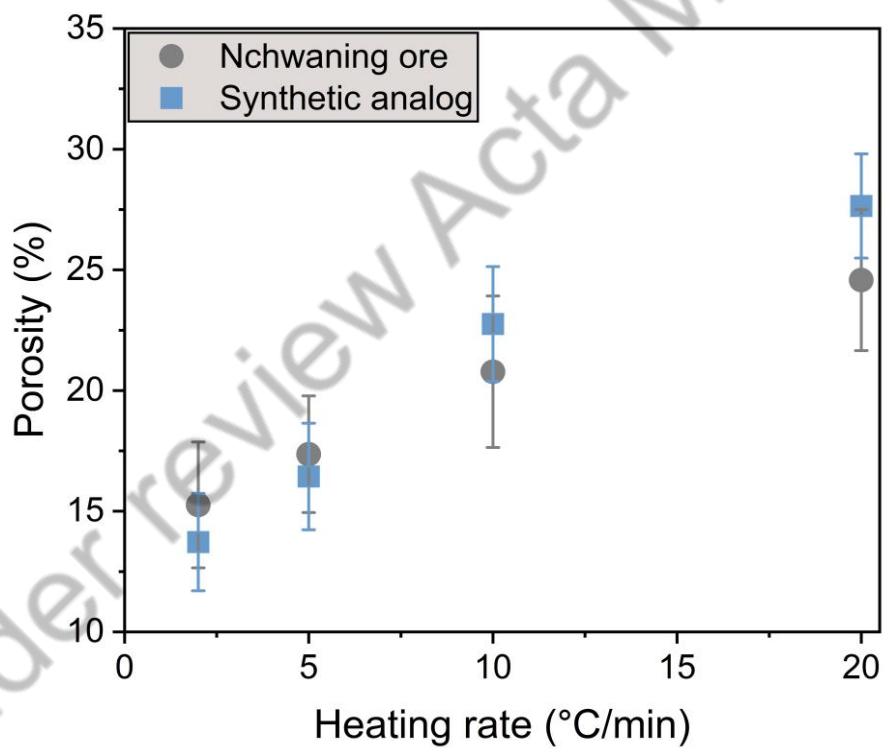

**Figure 7** - Final porosity (%) of hydrogen-reduced Nchwaning ore and synthetic analog samples as a function of heating rate (2–20 °C/min), measured after reduction to 900 °C. Porosity was quantified from SEM image analysis (n = 5 per condition and at 2000× magnification) using equivalent-area segmentation.

EBSD analysis was further employed to investigate the local microstructural differences of the synthetic samples reduced at 5 and 20 °C/min, Figure 8. Two phases were identified in all cases, namely, MnO and α-Fe. In the slow-heating case (5 °C/min, Figure 8a), the MnO grains are equiaxed and form a porous matrix with interspersed Fe grains. In addition, there are many dense spherical MnO grain clusters (with relatively larger grains) that do not contain any Fe particles. Overall, grain boundary lengths are moderate, and high-angle grain boundaries are prevalent (marked black lines in the EBSD maps). The inverse pole figure (IPF) map of the slow-heated sample exhibits no observable texture.

In contrast, the sample reduced at a fast heating rate of 20 °C/min exhibits a heterogeneous two-fold morphology. In Region I (Figure 8b), the microstructure is fine and a few small Fe particles (sub-micron) are distributed within the MnO matrix with irregularly shaped grains. The irregular-shaped MnO grains contain significant rotations and subgrains. The Fe particles often appear along grain boundaries or triple junctions of MnO. In this fine region, the MnO grain sizes are smaller than those in the sample reduced at a low heating rate of 5 °C/min, suggesting that rapid heating suppressed coarsening of the oxide matrix. In Region II (Figure 8c), large agglomerates stand out. These features consist of coarse MnO and Fe grains on the order of tens of microns. The phase map reveals that these large Fe agglomerates often envelop some MnO grains. Many grain boundaries here are serrated or irregular, and the color variation in the IPF maps suggests that adjacent grains often exhibit local texture (see encircled regions in Figure 8b, c), suggesting that they may have grown from the same parent orientation from larger oxide grains that formed during the rapid reduction. Further, the largest and most misshapen MnO grains are found adjacent to the large Fe particles. Despite the detection of $Mn_3O_4$ by XRD, it was not observed by EBSD; a discrepancy that may indicate that this phase consists of nanoscale precipitates that are below the resolution limit or that they are found locally in other regions of the sample.

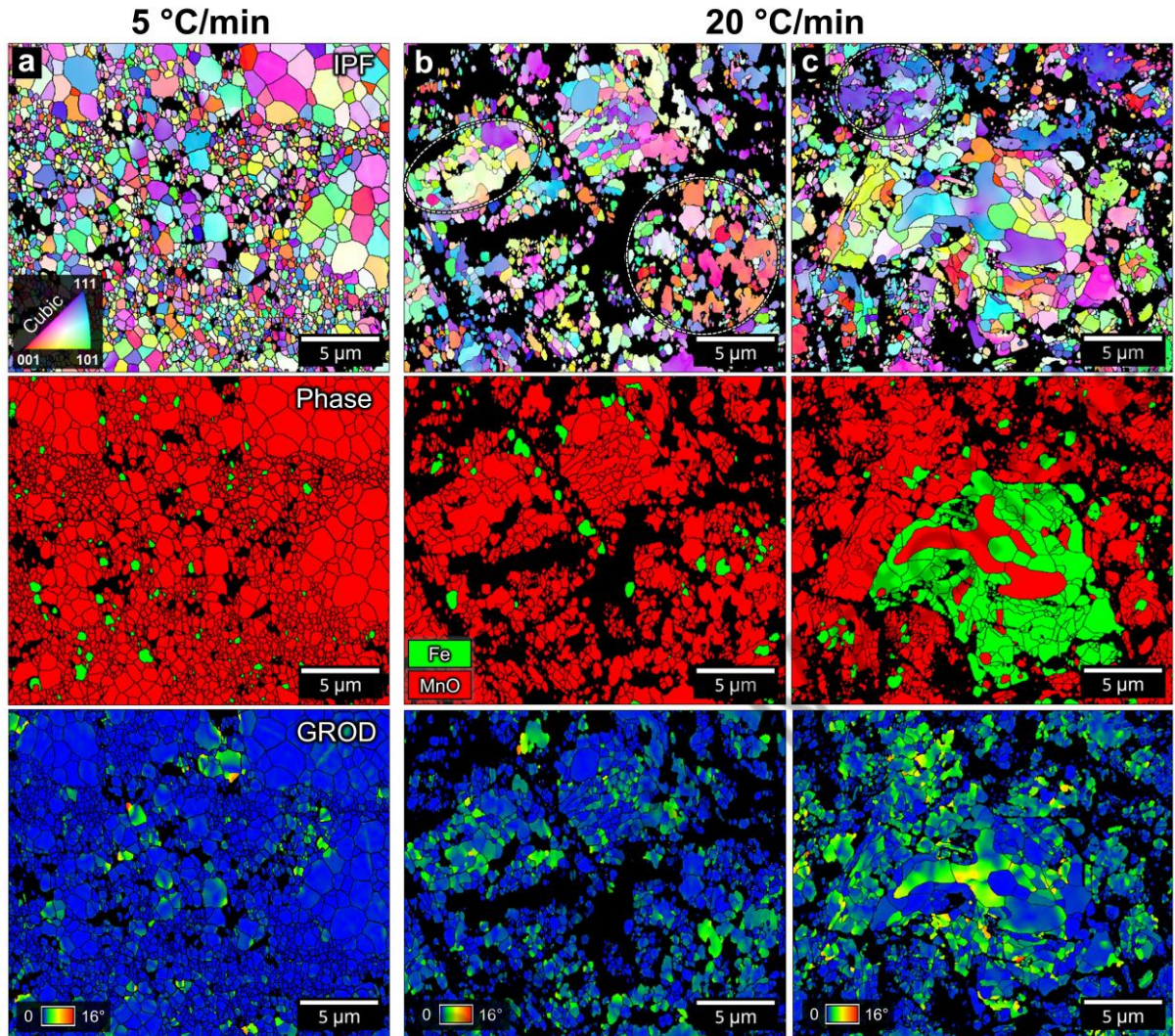

**Figure 8 -** EBSD analysis of $Mn_2O_3$-$9Fe_2O_3$ (wt.%) synthetic sample reduced at 900°C; Inverse pole figure (IPF, top row), phase (center row), and grain reference orientation deviation (GROD, bottom row) maps for the samples reduced at a heating rate of (a) 5 °C/min and two regions in the sample reduced at a heating rate of 20 °C/min: (b) area with fine Fe particulates and (c) area with large Fe grain agglomerates.

The grain reference orientation deviation (GROD) maps (Figure 8, bottom row) further highlight the impact of heating rate on residual lattice strain. In the sample reduced at 5 °C/min, grain-to-grain orientation deviations are generally low besides a few sporadic strained grains. High-angle grain boundaries dominate, and low-angle subgrains are rare. By contrast, the sample at the fast heating rate exhibits pronounced strain localization, especially inside the irregularly shaped grains. Notably, the highest rotation deviations occur around the Fe-rich agglomerates (Figure 8c). The in-grain rotations (subgrains) are found predominantly in the MnO phase and not in the metallic Fe; this is obvious when observed in the regions close to the larger Fe grains. This result implies that nucleation of metallic Fe and exsolution out of the MnO matrix generate localized strain and defects in those MnO grains.

## 4. Discussion

### 4.1 Mechanistic insights into multistep reactions

The kinetic analysis indicates that HyDR of ferromanganese oxides proceeds through multiple reaction steps with changing rate controlling mechanisms. In agreement with established pathways for Mn–Fe oxides, the ore and synthetic samples undergo the following sequence of reactions upon heating in $H_2$ atmosphere:

- Manganese oxides: $Mn_2O_3$ (bixbyite, $Mn^{3+}$) → $Mn_3O_4$ (hausmannite, mixed $Mn^{2+}/Mn^{3+}$) → MnO (manganosite, $Mn^{2+}$). This sequential reduction of Mn-oxides is well documented in the literature [35].
- Iron oxides: $Fe_2O_3$ (hematite) → $Fe_3O_4$ (magnetite) → FeO (wüstite, above 570 °C [36]) → Fe (metallic Fe) [14]. Under strongly reducing conditions with continuous heating, $Fe_3O_4$ can convert directly to Fe metal below 570 °C.

Each sub-reaction has its intrinsic kinetics, and the overall reduction behavior is a convolution of them. Initially, the reduction is controlled by chemical reaction, whereas at later stages it becomes controlled by solid-state diffusion. This transition is evident from the kinetic analysis and the $E_a$ is relatively low in early conversion (∼15-30 kJ/mol), but much higher at later conversion (∼50-70 kJ/mol), which is a hallmark of a change in rate-controlling step [37]. In the beginning, abundant free oxide surface is available for reaction with $H_2$, so the rate is governed by chemical reaction at the interface (nucleation and growth of product). Such processes typically have moderate activation energies (tens of kJ/mol) and can often be described by nucleation-growth models (*e.g.*, Avrami-Erofe'ev kinetics) [32]. Indeed, the sharp rise in reduction rate at the onset suggests a surface-controlled nucleation/growth reaction. As reduction proceeds, however, a solid product layer ($Mn_3O_4$/MnO or Fe) builds up around the unreacted core of each particle/grain (Figure 7). The system then shifts to diffusion-controlled kinetics, where oxygen atoms must diffuse out through the growing product layers to reach the surface and react with $H_2$ [35]. Diffusion through solids (or across phase boundaries) generally requires a higher activation energy and usually a lower reaction rate than surface chemical reaction [38]. Barner and Mantell's classic study on HyDR of $MnO_2$ noted that the formation of $Mn_3O_4$ made the reaction "progressively more protective," *i.e.*, diffusion-limited, as reduction proceeded [10].

The difference in initial sample microstructure and heterogeneity also affects the reduction process. The Nchwaning Mn ore is a natural material with inherent variations in composition, grain size, and porosity, whereas the synthetic sample is a homogeneous powder mixture of two pure oxides. This difference is manifested in the kinetic profiles: the ore reduction events (peaks in d$\alpha$/d$t$–$T$ profiles) are broader and less sharply defined (Figure 2a, b) compared with the more distinct d$\alpha$/d$t$ peaks in the synthetic samples (Figure 2c, d). In the ore, different particles (or different regions of a single particle) may reduce at slightly different temperatures due to variation in size, phase chemistry, and diffusion path lengths. This spreads out the reduction over a wider temperature range. Further, at low conversion degrees, the synthetic compact exhibits lower $E_a$ values than the ore, due to its finer microstructure, higher surface area, and shorter diffusion paths, which facilitate rapid surface-controlled reduction. However, as reduction progresses and the synthetic compact begins to densify, its $E_a$ rises more sharply than that of the ore, eventually surpassing it at high $\alpha$. This reversal is also evident in the peak-based Kissinger analysis (Figures 3b and d), where the synthetic sample shows lower $E_a$ for early peaks (1 and 2), but higher $E_a$ for peaks 3 and 4. These results imply that while the synthetic sample enables faster initial reduction, it becomes more susceptible to sintering-induced transport limitations at later stages, whereas the ore retains relatively accessible reaction fronts due to its stable porous morphology. Besides, the ore also contains impurity phases (gangue) such as silica or carbonates, which do not reduce but can react with Mn and Fe oxides. Silica, in particular, can form Mn silicates (*e.g.*, braunite) that consume Mn oxide and impedes its reduction [21, 39]. Note that there is a small portion of braunite (($Mn^{2+}Mn^{3+})_6(SiO_4)O_8$) in the ore sample (~2.6±1.8 wt.%) reduced with a heating rate of 10 °C/min (Figure 4a). In contrast, the synthetic oxide has no such inert constituents, which essentially behaves as an idealized ore. This model system is more readily reduced. Overall, the compositional and morphological heterogeneity of the ore leads to overlapping and extended reduction features, whereas the synthetic sample's homogeneity yields relatively better-separated reaction steps. Despite these differences, it is encouraging that the fundamental trend of increasing $E_a(\alpha)$ and the necessity of a multistep mechanism hold true for both materials, suggesting that the intrinsic kinetics of the Mn–Fe oxide system, rather than impurities or morphological differences of initial sample (Figure 1), dictate the overall reduction behavior.

## *4.2 Phase evolution and thermodynamics analysis*

The heating-rate-dependent TGA (Figure 2) and corresponding XRD results for the synthetic samples (Figure 5) show a clear trend that the total conversion degree decreases with an increase in heating rate. To investigate the phase transformation pathways during HyDR, interrupted reduction experiments were conducted at two heating rates, namely 5 °C/min and 20 °C/min, for synthetic analog samples. XRD was used to analyze the phase evolution at selected intermediate temperatures (Figure 9) for the distinct kinetic regimes identified as peaks in the TGA reduction rate curves (Figure 2d). These tests were interrupted at different temperatures corresponding to distinct reduction rate peaks; however, the total conversion degrees at each stage fall within a comparable range (within ±5%) between the two heating rates (see *α* values in Figure 8), enabling phase evolution to be evaluated at similar extents of reduction rather than just matching temperature.

The quantitative evaluation of phase fractions at all intermediate temperatures for samples reduced at different heating rates is provided in Figure 10a. The sample reduced at a slow heating rate of 5 °C/min consists of ~90.0 wt.% $Mn_3O_4$ (hausmannite spinel) and ~10.0 wt.% $Fe_2O_3$ (hematite) at 290 °C. The results confirm that at the slow heating rate the initial reduction is exclusively the Mn oxide step ($Mn_2O_3 \rightarrow Mn_3O_4$). In contrast, the $Fe_2O_3$ (hematite) phase remains largely present (9.9 wt.%). The onset of Fe oxide reduction is reached at 380 °C. In addition, nearly all $Mn_3O_4$ is reduced to MnO. Some FeO (wüstite) is present (3.1 wt.%), as distinct shoulders of the main MnO reflections, identifiable by the smaller lattice parameter of FeO (~4.33 Å) [40] compared with MnO (~4.44 Å) [41]. Given the low reduction temperature (here, 380 °C, at which the FeO is not thermodynamically stable), MnO halite may stabilize Fe-rich FeO as (Mn, Fe)O. This corresponds to the small Fe-oxide shoulder (the second peak) observed in the kinetic curves (Figure 2d). As the sample reaches 560 °C (the third peak), the major event corresponds to the reduction of FeO to α-Fe, accompanied by the reduction of $Mn_3O_4$ to MnO. Finally, at 660 °C, the reduction is halted with 92.3 wt.% MnO and 7.7 wt.% Fe in the reduced sample, corresponding to the complete reduction of $Mn_2O_3$ to MnO and $Fe_2O_3$ to α-Fe under given thermodynamic boundary conditions.

A different phase evolution was observed for the sample reduced at a high heating rate of 20 °C/min. For the first peak, at 390 °C, both Mn and Fe oxides began to reduce. Specifically, $Fe_3O_4$ (magnetite) was detected (3.7 wt.%) alongside MnO (86.8 wt.%) and a small amount of $Mn_3O_4$ (5.9 wt.%). This contrasts with the case at 5 °C/min, where magnetite did not form at an early reduction stage. At 470 °C, the emergence of Fe was observed alongside some FeO (2.5 wt.%), and a small amount of $Mn_3O_4$ (3.8 wt.%) is retained. Further, by the third kinetic event at 640 °C, it was observed that the FeO is dissolving into MnO (the FeO peaks merge with the MnO peaks). Finally, at 730 °C, the XRD pattern shows 6.46 wt.% metallic Fe and 90.12 wt.% MnO as the predominant phases, mirroring the reduced sample at a lower ramp rate, with miniscule amounts of retained $Mn_3O_4$ (2.76 wt.%). Note that there is a shift in the halite XRD reflections to higher angles and less α-Fe (6.0 wt.%), indicating the dissolution of Fe into MnO.

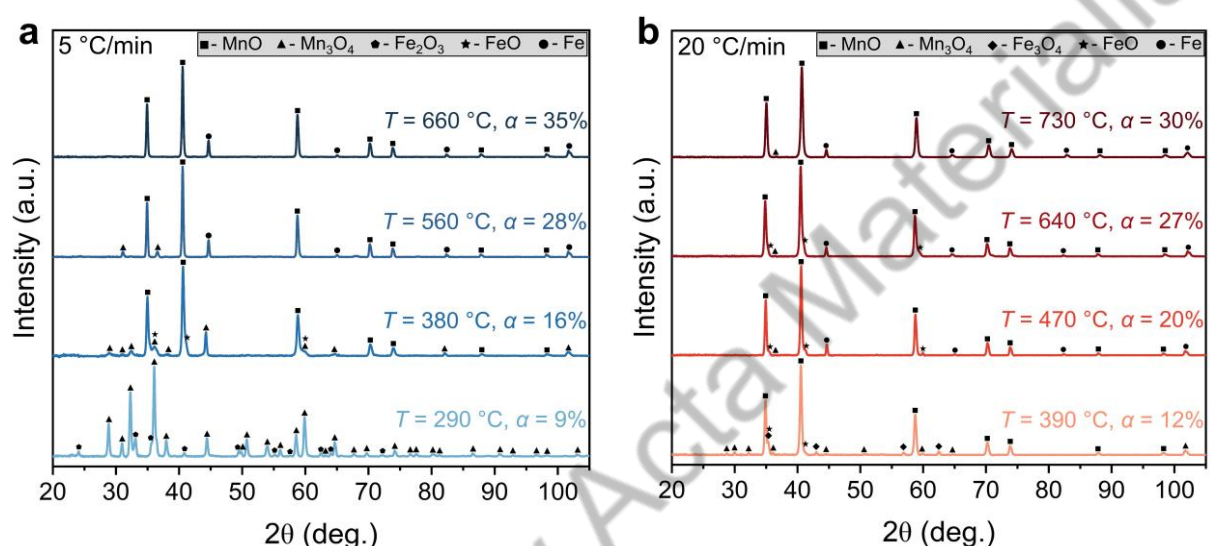

**Figure 9 -** Evolution of phase composition during hydrogen-based direct reduction of $Mn_2O_3$-9$Fe_2O_3$ (wt.%) synthetic analog sample at different interrupted temperatures for two heating rates: (a) 5 °C/min and (b) 20 °C/min. The selected temperatures correspond to the peaks presented in the reduction rate in Figure 2(d), and the conversion degree (α) for each interrupted test is indicated.

While the final products (*i.e.*, MnO and α-Fe) is the same for both heating rates, the reduction paths are different, as suggested by the interrupted XRD analysis. Generally, a slow heating regime is expected to allow the materials to follow the thermodynamic equilibrium phase transition pathways, while increasing the heating rate leads to divergence from equilibrium. Therefore, thermodynamic calculations are vital to elucidate the phase evolution under equilibrium conditions (Figure 10b). To gain deeper insights regarding the reduction behavior of Mn- and Fe-oxides, the Mn and Fe mole fractions in each phase are presented in Figure 10c, d, respectively. We chose to present the data for 600°C as it was shown by the interrupted XRD analysis to have accommodated the majority of phase transformations. Further data for 400°C (well below the stability of wüstite) is provided in Figure S8.

We first address the aspects of phase evolution involving Fe-oxides. Despite $Fe_2O_3 \rightarrow Fe_3O_4$ being regarded as a relatively easy reduction step [33], the first transformation in the slow heating rate regime is related to the $Mn_2O_3 \rightarrow Mn_3O_4$ reaction. This is easily explained by the higher thermodynamic stability of $Fe_2O_3$, as suggested by its phase region that is comparable to that of $Mn_3O_4$ in Figures 10e and 10f. The formation of $Fe_3O_4$ (magnetite) in the case of fast heating (and its absence in the case of slow heating) can be understood by both thermodynamic equilibrium and kinetic factors. Thermodynamically, $Fe_3O_4$ is the first lower valence state Fe-oxide product during reduction [38]. At slow heating, the sample spends longer time at intermediate temperatures, allowing nascent $Fe_3O_4$ to fully convert into Fe. In contrast, when heating fast, intermediate magnetite is retained as the $Fe_2O_3 \rightarrow Fe_3O_4$ step is relatively fast (lower activation barrier) compared with the $Fe_3O_4 \rightarrow Fe$ reaction [42]. Kinetically, the $Fe_2O_3 \rightarrow Fe_3O_4$ transformation is rapid, as it only requires an influx of rapid Fe cation diffusion, accommodated by electron flux. In contrast, the $Fe_3O_4 \rightarrow Fe$ transformations involve nucleation and growth of Fe and require relatively long-range oxygen diffusion (through the outer layer of Fe). At slow heating rate, the slower $Fe_3O_4 \rightarrow Fe$ reaction can keep pace, while for fast heating, the quicker chemistry of forming $Fe_3O_4$ outpaces Fe nucleation and growth [33]. Thus, the Fe-oxide early-stage reduction ($Fe_2O_3 \rightarrow Fe_3O_4$) is more nucleation-limited and diffusion-limited at slow and fast heating rates, respectively.

Lastly, it is evident that for both heating rate regimes, FeO exists at temperatures below 500 °C. This would not occur in the case of the single Fe-O system [43], but it can be explained by stabilizing wüstite as part of a halite solid solution with some fraction of MnO. According to thermodynamic calculations for 600 °C (Figure 10c, d), the mixed halite phase should contain about 5 at.% Fe and 45 at.% Mn (*i.e.*, $Fe_{0.1}Mn_{0.9}O$). Moreover, we show that the same is observed for a temperature of 400 °C (Figure S8), where halite would not exist for a pure Fe-O system. Since an obvious shoulder is observed for the main halite peaks in the XRD patterns (Figure 9), it can be postulated that there are two types of halite phases, one that is pure MnO (high intensity reflections) and the other mixed halite containing Fe (shoulder).

There are a couple of aspects related to Mn-oxides that are worth mentioning. For the initial stage of reduction (up to 2-3% conversion), $Mn_3O_4$ was the predominant phase for the slow heating, while it was already MnO in the case of fast heating. Furthermore, the fast heating resulted in the formation of more solid solution halite $(Mn_{1-x}Fe_x)O$ that persisted up to relatively high temperature (730 °C), as can be seen by the shift of the halite reflections to higher angles (noticeable for the high-angle peaks at around 100° 2θ). It is well known that (Mn,Fe)-mixed oxide phases can form at elevated temperatures enabled by cation mobility [44]. This can have a noticeable impact on reduction kinetics, for example, the presence of a few percent of $Mn_3O_4$ was enough to considerably lower the conversion rate of Fe oxides to Fe [45]. The differential oxygen mobility may also play a role; $Mn_3O_4 \rightarrow MnO$ is known to be diffusion-controlled with a relatively high activation energy (60-110 kJ/mol) [10], whereas $Fe_2O_3 \rightarrow Fe_3O_4$ (and $FeO \rightarrow Fe$) has a lower activation energy (50-80 kJ/mol) [33]. Thus, the heating rate can influence the phase evolution and prevalence of solid solution oxide formation, which could be one of the key factors affecting the reduction process kinetics and limiting the total conversion at higher heating rates.

The progression of $Mn_3O_4$ to MnO, likewise, highlights kinetic contrasts for the two heating rates. $Mn_3O_4$ formation is an initial, very quick step (surface reaction releasing O), whereas converting $Mn_3O_4$ to MnO requires breaking more bonds and diffusing oxygen out (forming $H_2O$). At 5 °C/min, once $Mn_3O_4$ is formed, the sample has ample dwell time as the temperature gradually rises through 400-600 °C (for 40 mins) for MnO nuclei to form and grow. At 20 °C/min, however, $Mn_3O_4$ once formed, is carried to a higher temperature before it completely converts to MnO. The faster heating provides less time for the solid-state reduction at intermediate temperatures. This aligns with the kinetic observation that the later stages of reduction have higher activation energies (~40 kJ/mol and above) and become diffusion-limited by product layers.

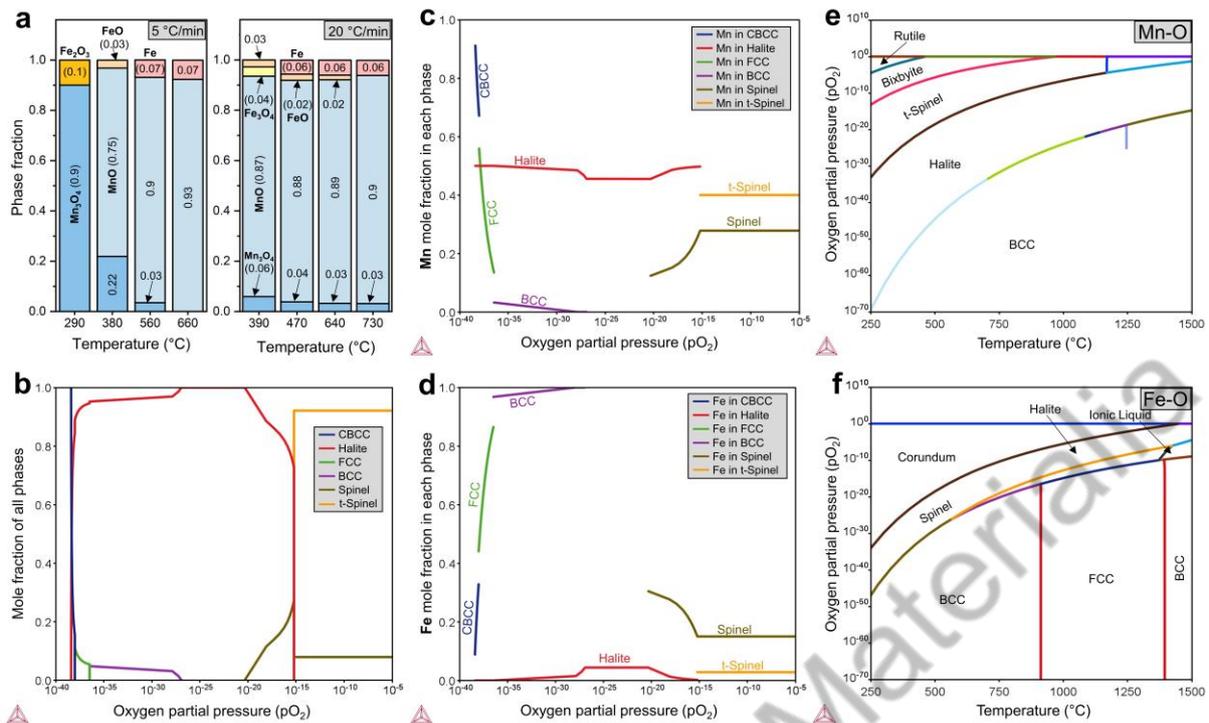

**Figure 10 -** (a) Experimentally determined phase fractions (by XRD) at different temperatures (from Figure 8) for two heating rates: 5 °C/min (left) and 20 °C/min (right). (b) Predicted phase stability diagram showing the mole fraction of all phases as a function of oxygen partial pressure ($pO_2$) at 600 °C. Calculated mole fractions of (c) Mn and (d) Fe mole in different phases as a function of $pO_2$ at 600 °C. CBCC refers to the α-Mn structure. Phase stability maps in the $p_{O2}$–temperature space for (e) Mn-O and (f) Fe-O systems. The maps illustrate the thermodynamic boundaries between different phases – for instance, $Fe_2O_3$ (corundum) is significantly more stable than $Mn_2O_3$ (bixbyite) and roughly the same as $Mn_3O_4$ (t-spinel).

## *4.3 Analysis of the microstructure evolution*

Given that the ferromanganese oxide system contains two distinct transition metal oxides, namely iron and manganese oxides, several key factors govern their reduction behavior and microstructure evolution. (1) Both Fe and Mn can form multiple oxide phases with different oxidation states ($M^{3+}$, $M^{3+/2+}$, and $M^{2+}$), each having different thermodynamic stability as a function of oxygen partial pressure. (2) Even for similar valence states, their crystal structures differ significantly: $Mn_2O_3$ (bixbyite, cubic) and $Mn_3O_4$ (spinel, tetragonal) differ from $Fe_2O_3$ (hematite, trigonal) and $Fe_3O_4$ (spinel, cubic), while both MnO and FeO adopt the cubic halite structure. Furthermore, spinel and halite-type phases may accommodate both cations in either ordered or disordered configurations [44]. (3) Finally, while Fe-oxides can be fully reduced to metallic Fe, MnO is thermodynamically stable under hydrogen and is not further reduced. These features enable a wide range of phase transformations and potential Fe-Mn oxide solid solutions during HyDR, particularly under non-isothermal conditions. In our case, the heating rate significantly influences the phase evolution sequence and transient reactions, in turn affecting the microstructure evolution. Nevertheless, the microstructural differences observed in Figures 6 and 8 can be rationalized by considering the thermodynamics, phase evolution, and kinetics at play during reduction at slow or fast heating rates.

In both the ore and synthetic samples, slow heating (2-5 °C/min) results in finer and more homogeneous microstructure (Figures 6a and 6c and S6a and S6b) whereas fast heating (10-20 °C/min) induces heterogeneity, including larger Fe agglomerates, denser MnO regions, and encapsulation phenomena (Figures 6b and 6d and S6c and S6d). The most prominent effect is the transformation in metallic Fe morphology: slow heating leads to uniformly dispersed, submicron spherical Fe particles (~50-100 nm to 1 μm), whereas fast heating yields a bimodal distribution with coarse Fe agglomerates (4-14 μm) alongside smaller particles (Figure 11a-d). These mechanistic interpretations are supported by prior high-temperature reduction studies. For instance, recent *in-situ* observations of hydrogen-reduced hematite pellets show that slow heating produces more porous Fe, whereas above a threshold rate, dense Fe layers appear and unreduced oxide cores persist [34]. Similarly, in exsolution phenomena in oxides, lower reduction temperatures (or slower ramping) lead to smaller, more numerous metal nanoparticles, while higher temperatures cause particle ripening and agglomeration [46].

Additionally, the MnO matrix shows morphological changes: after slow heating, it displays uniformly equiaxed grains (Figure 11e), whereas fast heating leads to irregularly shaped MnO domains (Figure 11f), further supporting the idea that reduction mechanisms and mass transport are more controlled under slower heating. When comparing the ore and synthetic samples, the synthetic analog consistently exhibits more homogeneous microstructures at all heating rates, with narrower particle and grain size distributions. This is attributed to its initially uniform composition and finer oxide powder precursors, as well as the absence of inert gangue phases or Mn-Si-O impurities that are present in the ore (e.g., braunite). These compositional and morphological differences also manifest in Fe-MnO partitioning and grain boundary development. The local texture in fast-reduced ore samples (Figure 8b–c) suggests that larger grains or spinel regions undergo grain refinement upon reduction, producing oriented clusters of MnO and Fe.

The development of porosity and Fe-MnO morphology is also strongly influenced by phase evolution and reduction kinetics. As shown in Figures 2 and 9, in slowly heated samples, the majority of $Fe_2O_3$ is reduced by ~600 °C (for the synthetic analog) or ~700-750 °C (for the ore), enabling formation of finely dispersed Fe and stable MnO. Conversely, in fast-heated samples, a significant fraction of the reduction (up to ~50% of the total mass loss) occurs above 700 °C. Hence, during slow heating, Fe-oxide reduces in a more controlled manner, and is practically reduced to Fe at intermediate temperatures of ~500-600 °C; while during fast heating, a considerable amount of Fe-oxide reacts with Mn-oxides (forming more thermodynamically stable mixed spinel and halite) before fully reducing to metallic Fe. As a result, some Fe remains in the oxide state up to high temperatures of >700 °C. The fast diffusion of Fe probably enables coalescence and coarsening of Fe grains as they exsolve at regions of Fe-rich halite phase. The local texture in the fast-heated samples (Figure 8b, c) also suggests that the reaction between oxides created larger grains that were then refined into textured grain clusters during the reduction process. Furthermore, the high-temperature exsolution of Fe out of (Fe,Mn)O is likely enabled by substantial defects which generate the shape irregularity and abundant subgrains in the resulting MnO matrix (Figure 8b, c). In addition, the dense Fe regions can encapsulate some Mn-oxide, thereby limiting outbound oxygen and being the reason that some residual $Mn_3O_4$ is detected by XRD in the samples reduced with fast heating rates (Figure 5).

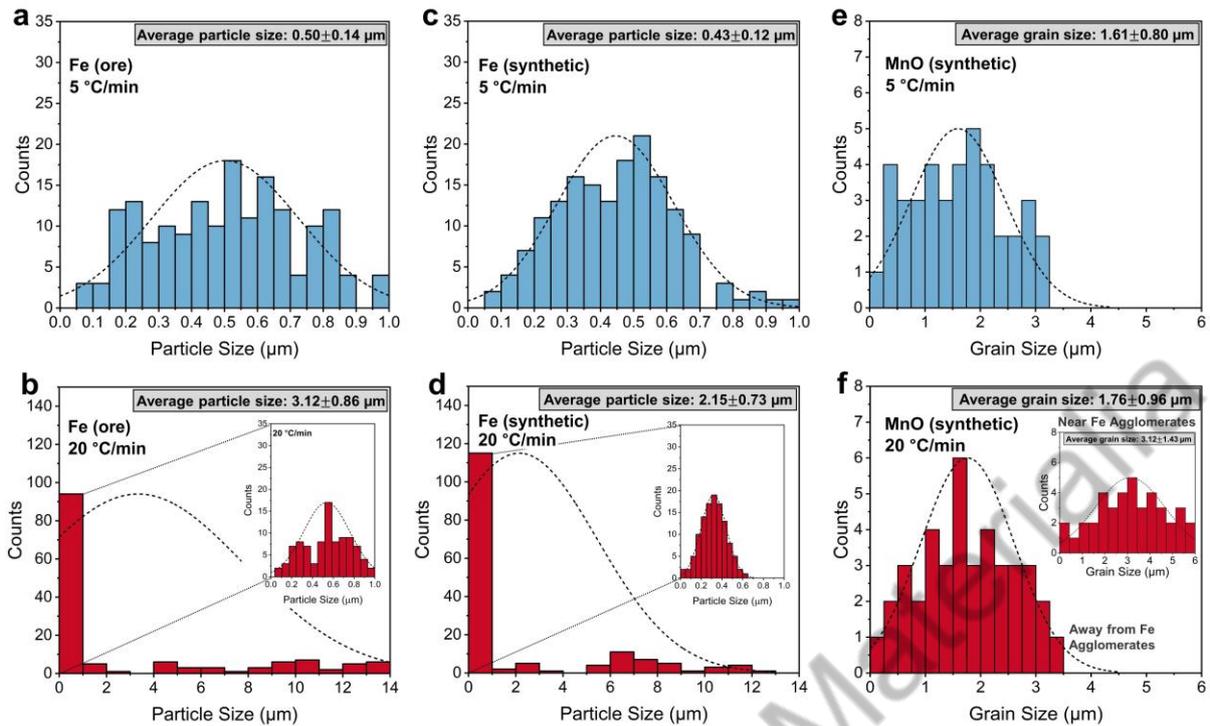

**Figure 11** - Histograms showing the particle size distributions of Fe particles in the hydrogen reduced samples prepared from (a, b) Nchwaning ore and (c, d) $Mn_2O_3$-$9Fe_2O_3$ (wt.%) synthetic sample subjected to heating rates of 5 and 20 °C/min, respectively. Insets in (b) and (d) show a magnified view of the submicron range (<1 μm) to reveal fine particles under the rapid heating condition. Histograms showing the grain size distributions of MnO in the hydrogen reduced synthetic samples for heating rates of (e) 5 and (f) 20 °C/min. Inset in (f) shows the grain size distributions of MnO near the large Fe agglomerates. Each subplot displays count-based histograms fitted with a Gaussian curve (dashed line) for visualization. The average particle/grain size and standard deviation for each condition are annotated in the inset boxes.

## *4.4 Impact of microstructure features on subsequent reduction processes*

The phase morphology from the pre-reduction step (Figures 6 and 8), particularly that of Fe, can significantly affect the subsequent downstream processing in both aluminothermic and carbothermic processes to produce metallic ferromanganese (Fe-Mn) alloys [47, 48].

The aluminothermic reaction is strongly exothermic ($3MnO + 2Al \rightarrow 3Mn + Al_2O_3$), leading to substantial heating up to temperatures that exceed the melting point of Fe-Mn (~1350 °C). Finer Fe particles have larger surface to volume ratios and act as thermal conduits and nucleation sites, promoting rapid formation of an Fe–Mn alloy as nascent Mn dissolves into the metallic Fe. The presence of fine Fe particles ensures uniformly distributed MnO, slightly lowers the activation energy and facilitates better heat transfer [49]. However, fine Fe dispersion can also present drawbacks. For instance, a previous study showed that metallic droplets <10 μm may remain suspended in the alumina-rich slag if not coalesced in time, potentially leading to metal losses [17].

In contrast, coarse Fe agglomerates heat and melt more slowly but form heavy, easily coalescing droplets that "seed" metal pools. These act as reservoirs for Mn incorporation, improving yield and promoting clean slag-metal separation [49]. While both fine and coarse Fe eventually dissolve into the Mn-rich alloy due to full Fe-Mn miscibility, the extent and timing of their participation differ. Fine Fe offers extensive Fe-Al interfacial area, promoting the formation of $FeAl_3$, which consumes more Al per Fe atom than $Fe_2Al_5$. This can locally deplete Al, limiting its availability for MnO reduction. Coarse Fe forms thin intermetallic layers relative to the volume, conserving Al for MnO reduction. Additionally, dissolved Mn stabilizes $FeAl_3$ at lower Al contents, exacerbating Al consumption when fine Fe dominates [50]. Hence, having finer Fe particles may lower Mn yield without sufficient excess of Al being provided.

In carbothermic reduction, both solid carbon and carbon monoxide (CO) can serve as reducing agents for MnO (MnO + C → Mn + CO). The presence of metallic Fe can play a catalytic role by facilitating these reduction reactions [51]. Fine Fe particles facilitate early nucleation of Fe-Mn-C alloys and accelerate reduction by dissolving carbon and reducing adjacent MnO. Thus, fine Fe distribution enhances MnO reduction kinetics, lowers the required temperature, and promotes uniform alloy formation [52]. The risk of metal entrapment in slag remains but is mitigated at typical smelting temperatures (1400-1600 °C) with sufficient stirring. Coarse Fe, however, does not contribute significantly to the reaction kinetics before melting begins. Once molten, it forms dense alloy pools that facilitate Mn absorption and promote clean phase separation [53]. Therefore, although still beneficial, coarse Fe is not as effective as fine Fe particles in facilitating carbothermic reduction.

In addition to phase morphology, the reduction-induced porosity generated during HyDR (~15-25 vol.%) plays a critical role in enabling efficient downstream metallothermic extraction. A higher fraction of coarser, interconnected pores improve access of molten reductants to unreduced MnO, potentially accelerating reaction kinetics and enhancing overall efficiency of aluminothermic reduction. However, the extent of this enhancement is governed by the percolation of the pore network and the wettability between molten Al and MnO surfaces [54, 55]. High porosity increases the melt–solid interfacial area, reducing kinetic barriers and facilitating local diffusion. It also promotes capillary-driven infiltration of molten Al or Fe–C alloys into the porous MnO matrix, thereby improving melt penetration and reactivity [47, 56]. The wettability of molten Al on MnO and Fe is strongly temperature-dependent but is significantly improved in porous systems due to triple-phase boundary formation and melt entrapment within pore channels. This promotes intimate Al–MnO contact and accelerates aluminothermic reduction. Likewise, in carbothermic reduction, a well-connected porous network facilitates the escape of CO gas and drainage of liquid alloy phases, reducing internal pressure buildup and enabling effective percolation of Fe-Mn-C melts [57].

Overall, slow heating yields a finely dispersed Fe–MnO microstructure with high surface area and interconnected porosity, which promotes fast, uniform reduction and efficient melt infiltration, but may require careful slag management to prevent metal entrapment. In contrast, fast heating leads to coarser Fe agglomerates and less uniform but still highly porous MnO domains, which simplify droplet coalescence and metal-slag separation, though at the cost of slightly elevated reduction temperatures or increased reductant demand.

## 5. Conclusions

This study provides a mechanistic understanding of the effect of heating rate on the hydrogen-based direct reduction behavior of ferromanganese oxides. Using a combination of thermogravimetric analysis, isoconversional kinetics assessment, interrupted microstructure analysis by XRD, SEM, EBSD and EDX, as well as thermodynamic calculations, the main conclusions are summarized as follows:

- Both the natural Nchwaning ore and the synthetic $Mn_2O_3$-$Fe_2O_3$ analog reduce via sequential phase transitions, governed by distinct kinetic regimes. Early-stage reductions are dominated by surface-reaction-controlled nucleation and growth, while later stages become solid-state diffusion-limited due to the formation of product layers, as evidenced by a systematic increase in activation energy with conversion.

- Slow heating rate (2-5 °C/min) promotes more sequential phase evolution during reduction and porous microstructure development, enabling complete transformation from $Mn_2O_3$ to MnO and $Fe_2O_3$ to Fe. In contrast, rapid heating (10-20 °C/min) thermally arrests the reduction pathway, resulting in the persistence of $Mn_3O_4$ and lower reduction extent due to insufficient time to complete reduction, compounded by the formation of dense Fe agglomerates that locally isolate Mn-oxide.

- Thermodynamic calculations and interrupted XRD analysis revealed that Mn stabilizes the spinel $(Mn,Fe)_3O_4$ and halite $(Mn,Fe)O$ phases, thus requiring lower oxygen partial pressures to reduce them. Fast heating promotes oxide intermixing, further hindering full conversion by stabilizing these mixed oxides.

- FeO formed at low temperatures (<400 °C), well below the thermodynamic stability of pure wüstite (570 °C) is likely due to the stabilization by Mn, as supported by thermodynamic calculations.

- Slow heating results in a more homogenous MnO matrix with finely dispersed submicron Fe particles and low internal strain. Conversely, fast heating produces microstructural heterogeneity, including coarse Fe agglomerates with Mn-oxide embedded in them, and more local texture and strain accumulation (in the form of subgrains) in the MnO matrix. These features are closely tied to the morphological evolution during Fe partitioning out of the mixed oxide phases at later stages of reduction.
- Porosity development is likewise influenced by heating rate. For both ore and synthetic analogs, total porosity increases with heating rate (~15% to 30%), This increase in porosity can have significant implications for subsequent metallothermic reduction processes, as it enhances melt infiltration and interfacial contact with MnO, potentially improving downstream efficiency.

The findings underscore how thermally modulated reduction pathways directly dictate phase formation and product morphology, factors critical for downstream extraction processes. By tailoring the hydrogen-based reduction conditions, it is possible to engineer microstructures that enhance the efficiency and selectivity of subsequent aluminothermic or carbothermic processing. This study provides mechanistic insights into phase evolution in complex oxide systems and establishes a foundation for thermally guided optimization of ferromanganese production.

## CRediT authorship contribution statement

**Anurag Bajpai:** Conceptualization, Methodology, Formal analysis, Investigation, Validation, Visualization, Writing – original draft; **Barak Ratzker:** Formal analysis, Investigation, Validation, Visualization, Writing – original draft; **Shiv Shankar:** Investigation, Visualization; **Dierk Raabe:** Resources, Writing – review and editing, Supervision; **Yan Ma:** Conceptualization, Resources, Funding acquisition, Writing – review and editing, Supervision.

## Declaration of Competing Interest

The authors declare that they have no known competing financial interests or personal relationships that could have appeared to influence the work reported in this paper.

**Acknowledgements**

A.B. and B.R. are grateful for the financial support of respective Alexander von Humboldt Foundation Fellowships (hosted by D.R.). A.B., S.S., and Y.M. acknowledge the financial support from the Horizon Europe project HAlMan, co-funded by the European Union grant agreement (ID 101091936). D.R. and Y.M. are grateful for the financial support through the ERC Advanced grant ROC (Grant Agreement No 101054368). Views and opinions expressed are, however, those of the author(s) only and do not necessarily reflect those of the European Union or the ERC. Neither the European Union nor the granting authority can be held responsible for them. The authors thank Benjamin Breitbach for his support of XRD experiments and Katja Angenendt and Christian Broß for their support of the metallography lab and SEM facilities.

# Supplementary Information

# Sustainable Pre-reduction of Ferromanganese Oxides with Hydrogen: Heating Rate-Dependent Reduction Pathways and Microstructure Evolution

A. Bajpai[1,*], B. Ratzker[1], S. Shankar[1], D. Raabe[1], Y. Ma[1,2*]

[1]Max Planck Institute for Sustainable Materials, Max-Planck-Str. 1, Düsseldorf, 40237 Germany

[2]Department of Materials Science and Engineering, Delft University of Technology, Mekelweg 2, 2628 CD Delft, the Netherlands

*Corresponding authors: a.bajpai@mpie.de; d.raabe@mpie.de; y.m.ma@tudelft.nl

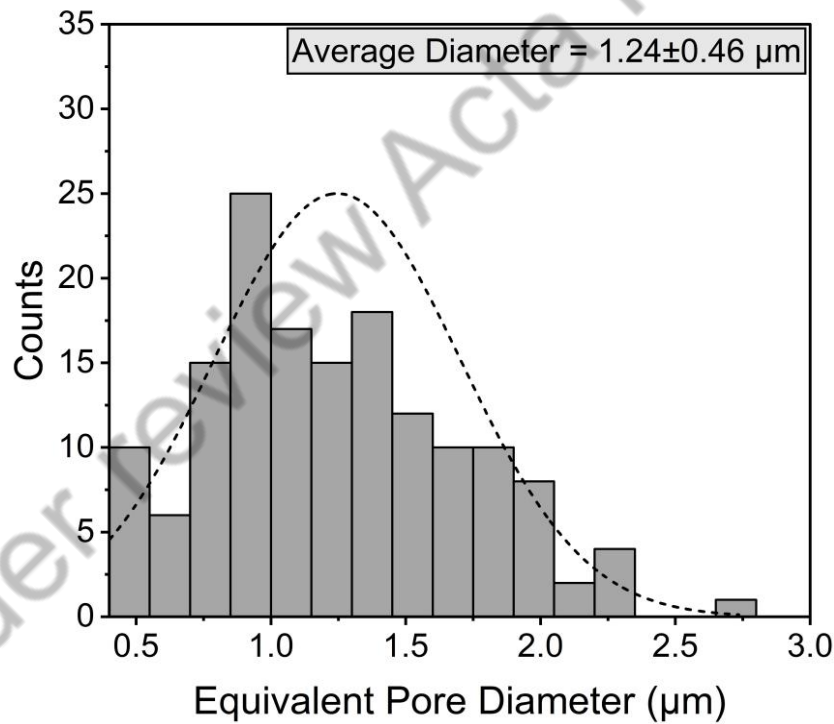

**Figure S1** - Distribution of the equivalent diameter of pores for the Nchwaning ore sample.

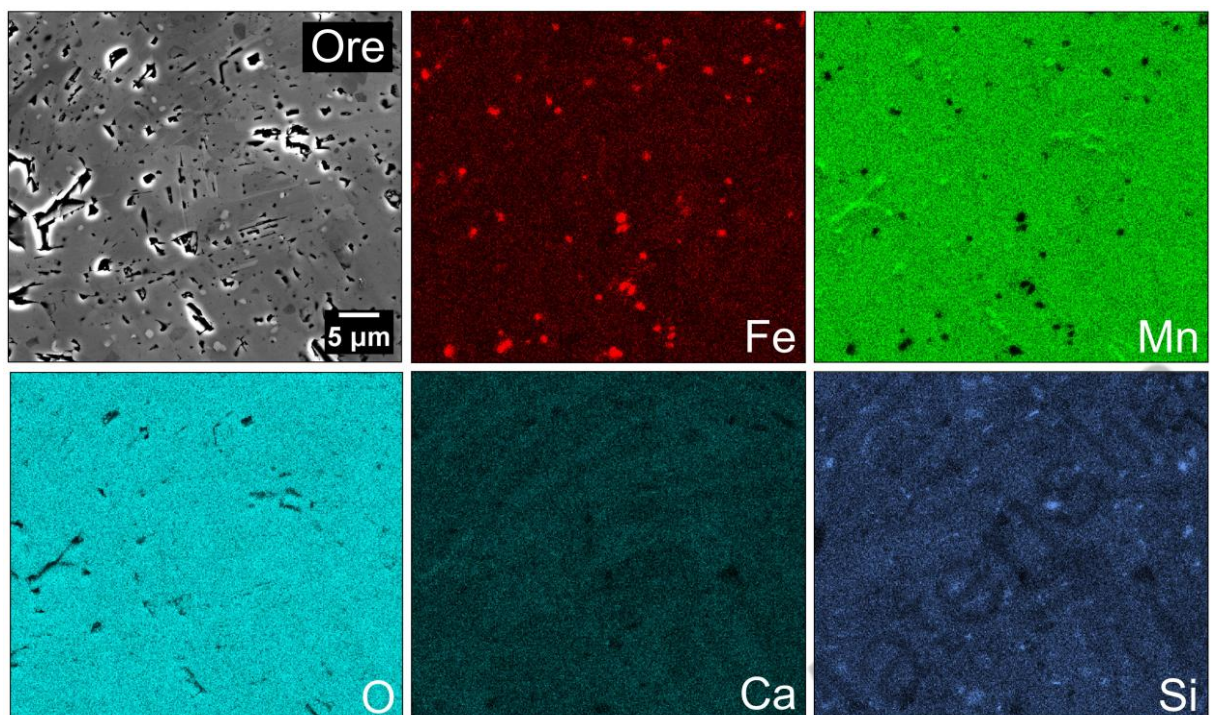

**Figure S2** - Scanning electron microscope (SEM) Backscattered electron (BSE) micrograph and corresponding energy-dispersive X-ray spectroscopy (EDS) elemental maps of the Nchwaning ore, highlighting the spatial distribution of key elemental constituents.

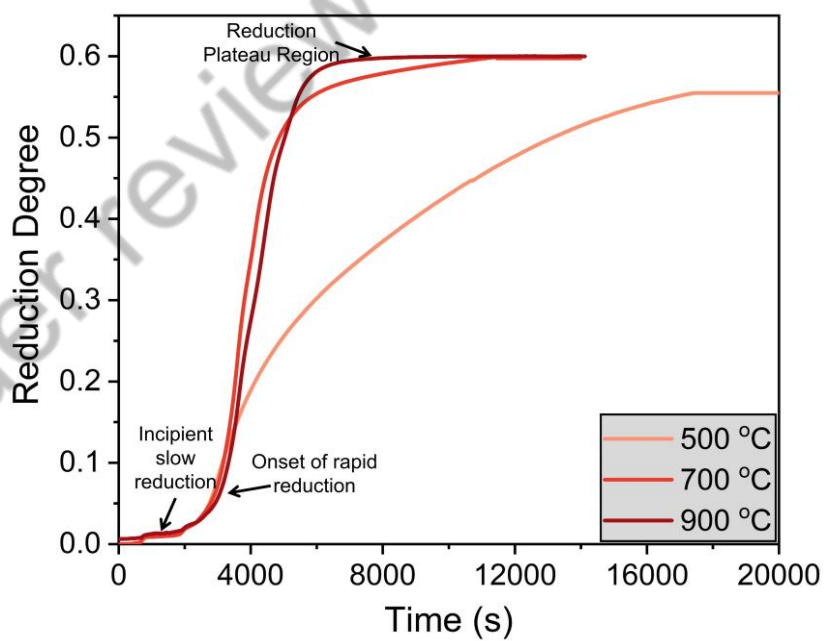

**Figure S3** - Reduction degree vs. time for the $Mn_2O_3$-$9Fe_2O_3$ (wt. %) synthetic sample under isothermal hydrogen reduction at 500 °C, 700 °C, and 900 °C. The curves exhibit three distinct kinetic regimes: an initial incipient slow reduction, a sharp onset of rapid reduction, and a final plateau region corresponding to reduction saturation. Higher temperatures accelerate the transition to the rapid reduction regime and lead to a faster attainment of the reduction plateau.

# Supplementary Note S1: Theory and Comparative Framework of Isoconversional Kinetic Methods

The solid-state reaction rate under non-isothermal conditions can be expressed by the following general equation:

$$\frac{d\alpha}{dt} = k(T)\,f(\alpha)$$

where, $\alpha$ is the fraction of conversion ($0 \leq \alpha \leq 1$), $t$ is time, $k(T)$ is the temperature-dependent rate constant, and $f(\alpha)$ is the reaction model, dependent on the reaction mechanism.

The temperature dependence of the rate constant $k(T)$ is typically modeled by the Arrhenius equation:

$$k(T) = k_0\,exp(-E_\alpha/RT)$$

where, $k_0$ is the pre-exponential factor, $E_\alpha$ is the apparent activation energy, $R$ is the universal gas constant, and $T$ is the absolute temperature.

Under non-isothermal conditions, where heating occurs at a constant rate $\beta = dT/dt$, the rate equation becomes;

$$\frac{d\alpha}{dT} = \left(\frac{1}{\beta}\right)\frac{d\alpha}{dt} = \left(\frac{k_0}{\beta}\right)exp(-E_\alpha/RT)\,f(\alpha)$$

Integration yields:

$$g(\alpha) = \int_0^\alpha \frac{d\alpha}{f(\alpha)} = \left(\frac{k_0}{\beta}\right)\int_{T_0}^T exp(-E_\alpha/RT)\,dT$$

Since $f(\alpha)$ is typically unknown for complex multi-step reactions, isoconversional methods bypass this by evaluating the activation energy directly from the temperature at which a constant α is reached across multiple heating rates.

The three isoconversional approaches used in this study differ in how they approximate or integrate the temperature integral and whether they rely on integral or differential formulations:

| Method | Equation | Type | Strengths | Limitations |
| --- | --- | --- | --- | --- |
| Kissinger–Akahira–Sunose (KAS) | $ln\left(\frac{\beta}{T^2}\right) = \ln\left(\frac{AR}{E_\alpha g(\alpha)}\right) - \frac{E_\alpha}{RT_\alpha}$ | Integral | More accurate at high T; broad kinetic range | Assumes constant $E_\alpha$ in α interval |
| Ozawa–Flynn–Wall (OFW) | $ln(\beta) = \ln\left(\frac{AE_\alpha}{Rg(\alpha)}\right) - 2.315 - \frac{0.4567 E_\alpha}{RT_\alpha}$ | Integral | No need for $T^2$ term; model-free | Slightly less accurate at low T |
| Friedman | $ln\left(\frac{d\alpha}{dt}\right) = \ln[k_0 f(\alpha)] - \frac{E_\alpha}{RT_\alpha}$ | Differential | Captures the mechanism changes | Sensitive to experimental noise |

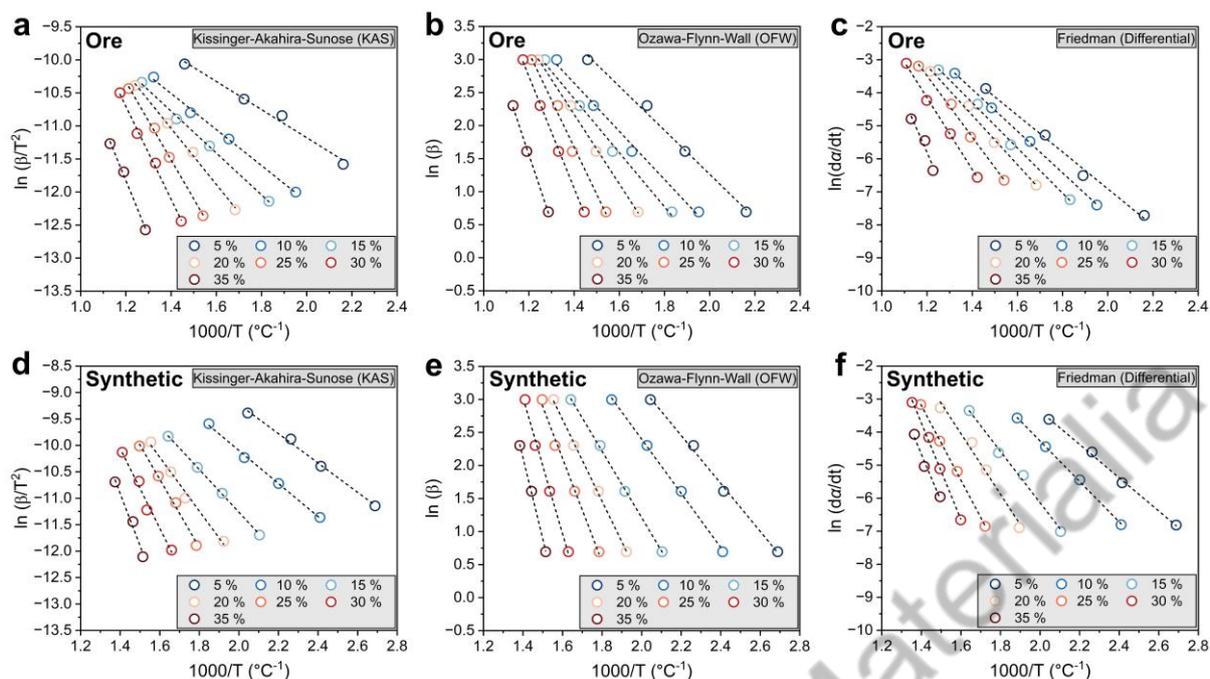

**Figure S4** - Isoconversional analysis of apparent activation energy for hydrogen-based reduction of (a–c) Nchwaning ore and (d–f) $Mn_2O_3$-$9Fe_2O_3$ (wt. %) synthetic sample using three model-free methods: (a, d) Kissinger–Akahira–Sunose (KAS), (b, e) Ozawa–Flynn–Wall (OFW), and (c, f) Friedman differential method. Each plot corresponds to different conversion levels (10%–50%) marked by colored data points. The linear trends validate the application of isoconversional approaches, and the extracted activation energies provide insight into the multistep nature of the reduction process.

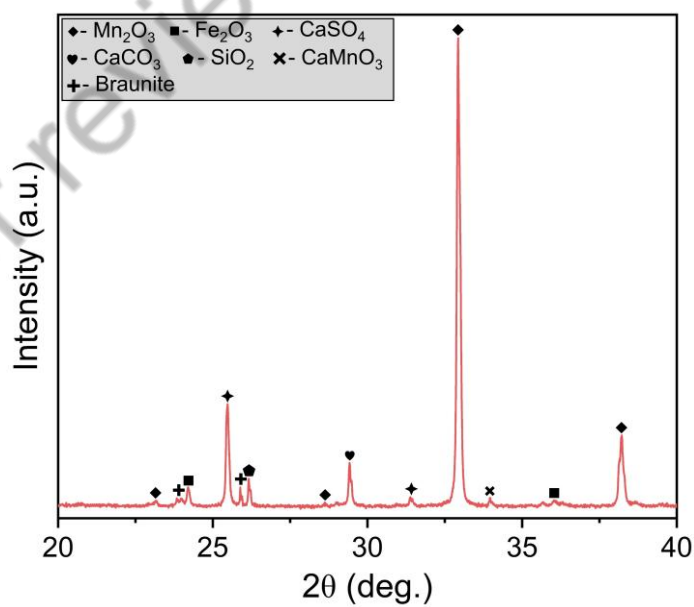

**Figure S5** - Enlarged view of the X-ray diffraction (XRD) pattern for the Nchwaning ore sample in the 2θ range of 20°–40°, highlighting the presence of minor gangue phases. This magnified view confirms the heterogeneous nature of the ore, which contains both iron and manganese oxides as well as various gangue phases.

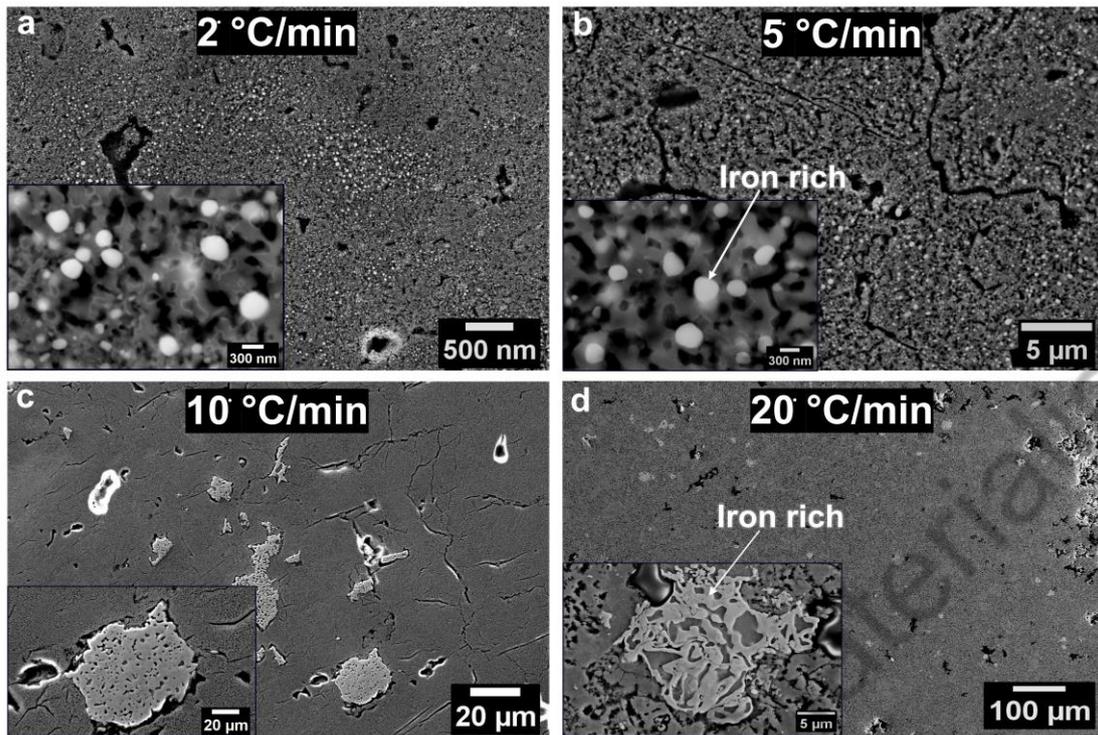

**Figure S6** - SEM micrographs showing the morphology of reduced Nchwaning ore as a function of heating rate during hydrogen-based direct reduction: (a) 2 °C/min, (b) 5 °C/min, (c) 10 °C/min, and (d) 20 °C/min. Insets provide higher-magnification views revealing the microstructural features of Fe precipitates. At lower heating rates (2–5 °C/min), a dense dispersion of nanoscale metallic Fe-rich precipitates is observed. As the heating rate increases (10–20 °C/min), large micron-scale agglomerates and interconnected Fe structures form, accompanied by pronounced coarsening and sintering.

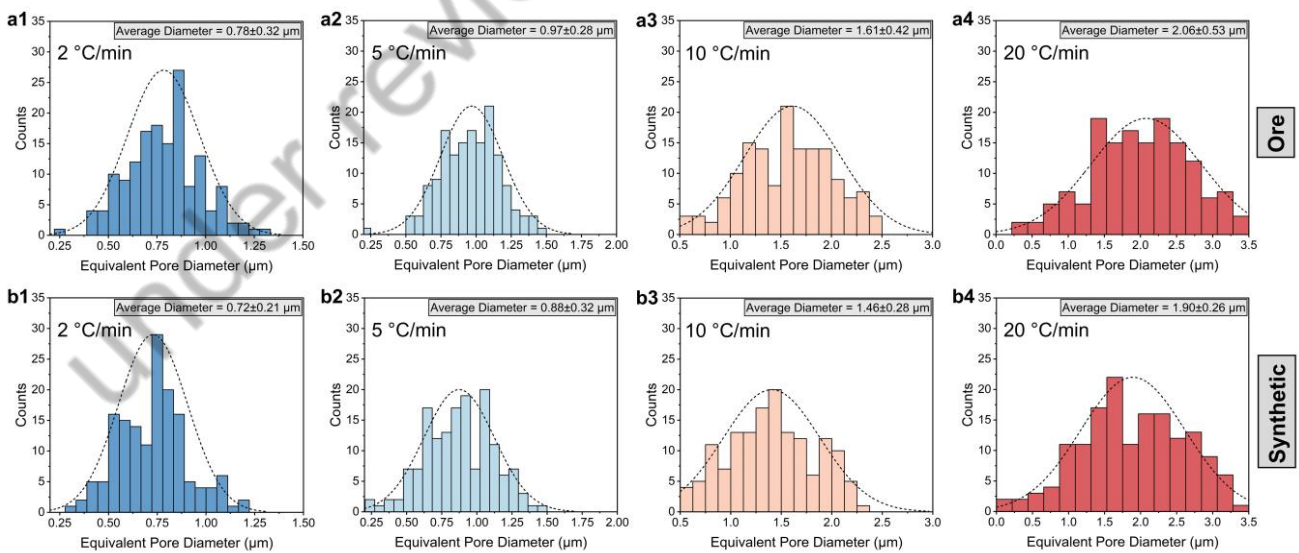

**Figure S7** - Pore size distributions of hydrogen-reduced Nchwaning ore (top row: a1–a4) and synthetic analog (bottom row: b1–b4) as a function of heating rate: 2 °C/min (a1, b1), 5 °C/min (a2, b2), 10 °C/min (a3, b3), and 20 °C/min (a4, b4). The dashed lines represent log-normal fits.

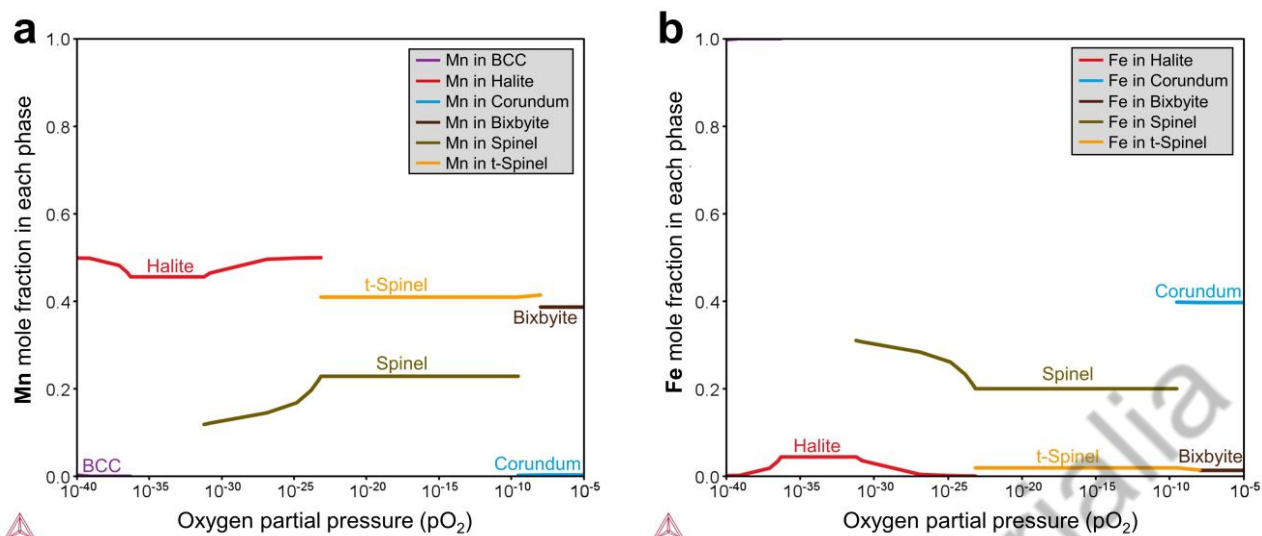

**Figure S8** - Calculated mole fractions of (a) Mn and (b) Fe in different phases as a function of $pO_2$ at 400 °C.

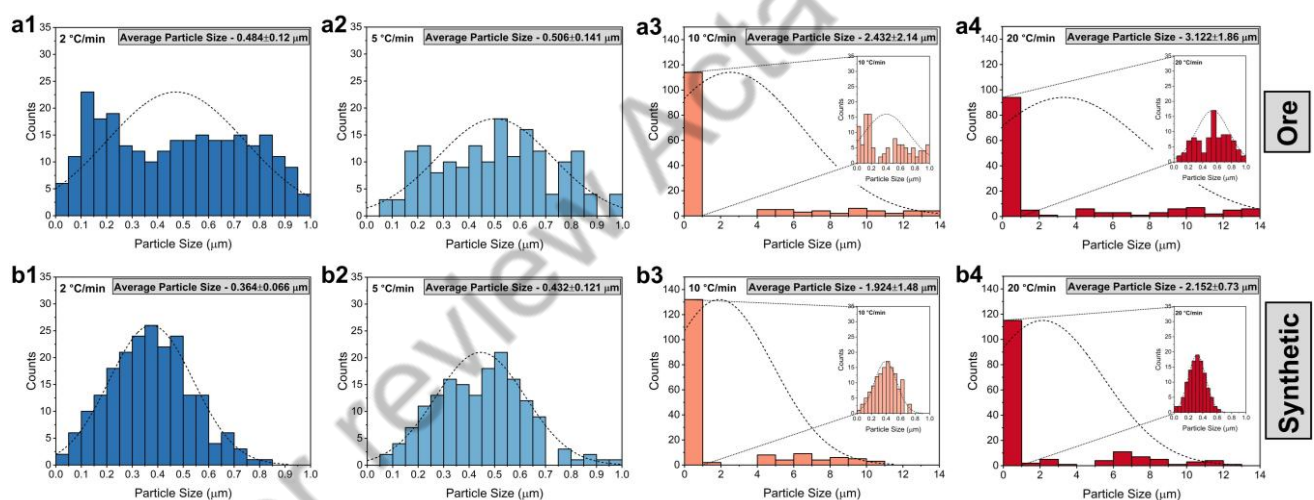

**Figure S9** - Histograms showing the particle size distributions of Fe in the hydrogen reduced samples prepared from Nchwaning ore (a1–a4) and $Mn_2O_3$-$9Fe_2O_3$ (wt.%) synthetic sample (b1–b4) subjected to heating rates of 2, 5, 10, and 20 °C/min, respectively. Each subplot displays count-based histograms fitted with a Gaussian curve (dashed line) for visualization. The average particle size and standard deviation for each condition are annotated in the inset boxes. Insets in (a3, a4, b3, b4) show a magnified view of the submicron range (<1 μm) to emphasize fine particle retention under rapid heating.

**Table S1** - Elemental composition (in wt.%) of the Nchwaning ore and its synthetic analog before and after hydrogen-based reduction, as determined by energy-dispersive X-ray spectroscopy (EDS). The values represent the mean ± standard deviation across at least 10 multiple measurements.

| Nchwaning Ore | | | Synthetic Analog Sample | | |
|---|---|---|---|---|---|
| **Element** | **Weight %** (Initial Sample) | **Weight %** (Reduced Sample) | **Element** | **Weight %** (Initial Sample) | **Weight %** (Reduced Sample) |
| O | 40.81 ± 2.92 | 30.24 ± 3.50 | O | 40.20 ± 3.23 | 31.12 ± 3.63 |
| Si | 2.10 ± 1.26 | 2.03 ± 1.58 | Si | -- | -- |
| Ca | 3.58 ± 1.11 | 3.43 ± 1.82 | Ca | -- | -- |
| Mn | 47.65 ± 3.04 | 56.60 ± 3.20 | Mn | 53.18 ± 3.07 | 60.34 ± 3.18 |
| Fe | 5.86 ± 1.48 | 7.69 ± 1.83 | Fe | 6.52 ± 1.36 | 8.56 ± 1.45 |

**Table S2** – The activation energy values (kJ/mol) for hydrogen-based reduction of natural Nchwaning ore and its synthetic analog, calculated using model-free isoconversional methods: Kissinger–Akahira–Sunose (KAS), Ozawa–Flynn–Wall (OFW), and Friedman differential method, across different degrees of conversion ($\alpha$).

| | **Reduction degree, $\alpha$ (%)** | **Activation Energy (kJ/mol)** | |
|---|---|---|---|
| | | **Nchwaning Ore** | **Synthetic Analog** |
| **KAS** | 5 | 17.69 ± 3.68 | 16.11 ± 3.90 |
| | 10 | 22.66 ± 3.52 | 20.49 ± 3.51 |
| | 15 | 28.83 ± 3.88 | 25.34 ± 3.88 |
| | 20 | 37.75 ± 4.07 | 30.75 ± 4.11 |
| | 25 | 49.01 ± 4.23 | 37.59 ± 4.37 |
| | 30 | 58.41 ± 4.21 | 43.75 ± 4.86 |
| | 35 | 69.97 ± 3.51 | 54.86 ± 4.63 |
| **OFW** | 5 | 27.68 ± 3.67 | 23.76 ± 3.94 |
| | 10 | 32.84 ± 3.81 | 28.20 ± 3.72 |
| | 15 | 39.91 ± 3.53 | 34.62 ± 4.14 |
| | 20 | 48.27 ± 3.92 | 42.10 ± 4.76 |
| | 25 | 59.27 ± 4.26 | 49.04 ± 4.14 |
| | 30 | 70.39 ± 3.92 | 56.61 ± 4.42 |
| | 35 | 85.05 ± 3.83 | 71.02 ± 6.36 |
| **Friedman (Differential)** | 5 | 46.33 ± 4.01 | 39.12 ± 3.14 |
| | 10 | 52.61 ± 3.82 | 47.58 ± 4.35 |
| | 15 | 58.66 ± 4.60 | 56.14 ± 4.52 |
| | 20 | 67.74 ± 5.13 | 63.55 ± 5.16 |
| | 25 | 76.84 ± 5.34 | 73.05 ± 4.12 |
| | 30 | 90.13 ± 5.05 | 82.29 ± 4.41 |
| | 35 | 105.33 ± 6.07 | 95.74 ± 5.68 |